\documentclass[preprint,
 amsmath,amssymb,
pra]{revtex4-1}

\usepackage{graphicx}
\usepackage{dcolumn}
\usepackage{bm}
\usepackage[mathlines]{lineno}
\usepackage{xr}
\usepackage{array}
\usepackage{hyperref,amsmath,amsthm,amssymb, graphicx, array, bbm, enumitem, physics}
\usepackage{cleveref}
\usepackage{lineno}
\usepackage{titlesec}
\usepackage{comment}
\usepackage{minted}

\newcommand{\Id}{\text{Id}}

\newcommand{\bx}{\mathbf{x}}
\newcommand{\bk}{\mathbf{k}}

\newcommand{\bp}{\mathbf{p}}
\newcommand{\by}{\mathbf{y}}

\newcommand{\be}{\mathbf{e}}
\newcommand{\hp}{\phi}

\newcommand{\hs}{\sigma}

\newcommand{\M}{\mathcal{M}}

\usepackage{float}
\usepackage{longtable}

\usepackage{physics}
\usepackage{quantikz}
\usepackage{color}

\begin{document}

\preprint{APS/123-QED}

\title{Effect of atomic interactions on quantum communication protocols}

\author{Zixiang Hong}
\email{tom.hong@yale.edu}
\affiliation{Department of Applied and Computational Mathematics, Yale University, New Haven, Connecticut 06511, USA}
 
\author{John C. Schotland}
\email{john.schotland@yale.edu}
\affiliation{%
Department of Mathematics and Department of Physics, Yale University, New Haven, Connecticut 06511, USA
}%

\date{\today}

\begin{abstract}
We develop a first-principles framework for analyzing photonic quantum communication in an atomic medium. Starting from a real-space quantization of the electromagnetic field coupled to an ensemble of two-level atoms, we derive an effective  model for polarization-encoded single-photon states. We apply this framework to superdense coding to study communication efficiency and to the BB84 quantum key distribution protocol for communication security. For superdense coding, we evaluate the protocol capacity of the fixed Bell-state alphabet under erasure, dephasing, and depolarizing channels and show that it decreases monotonically with increasing atom-field coupling. For BB84, we formulate a modified protocol that discards erasure outcomes and uses the microscopic model to estimate the intrinsic loss rate. We derive the corresponding asymptotic key rate and show that it decreases monotonically with coupling strength and vanishes beyond a finite critical coupling.
\end{abstract}

\maketitle


\section{Introduction}
Quantum technology holds the promise of revolutionizing communication by offering enhanced efficiency and security \cite{Nielsen_Chuang_2010}. Protocols such as superdense coding, for example, leverage entanglement to enable the transmission of two classical bits of information while sending only a single qubit \cite{Bennett92}. In the realm of security, quantum key distribution (QKD) protocols provide a mechanism to detect potential eavesdroppers, thereby ensuring the confidentiality of the transmitted information \cite{BEN84}. The feasibility of these protocols has been confirmed through successful experimental demonstrations of both superdense coding \cite{Barreiro2008,Williams17,Hu18} and BB84 \cite{Rosenberg07,Liao18,Sibson2017}. Photonic approaches are said to be the leading candidates for implementing such communication tasks \cite{Knill2001}. 

Photons are a fragile resource, susceptible to losses that degrade the quality of communication \cite{Gisin02,Scarani09}. These losses can be broadly categorized as intrinsic, arising from interactions with the transmission medium, and external, stemming from factors like leakage, thermal effects, or measurement errors. A recent publication \cite{hong2026onephotoncommunicationatomicmedia} has underscored the significance of these interactions, revealing that the effects of an atomic medium during transmission impose a fundamental lower bound on transmission fidelity. A rigorous analysis of these loss mechanisms is therefore crucial for developing future large-scale quantum networks. However, previous research in this area has often either neglected atomic effects or relied on phenomenological models \cite{Buttler98,Pati05,Zaunders26} and classical scattering theories \cite{Fasel2004,Galtarossa2005,GISIN1992}, rather than a first-principles quantum approach.

To address this gap, we develop a real-space quantization formalism that incorporates polarization and builds on the first-principles quantum electrodynamics framework introduced in \cite{hong2026onephotoncommunicationatomicmedia}. This approach systematically accounts for interactions with a two-level atomic medium as well as extrinsic noise processes. We apply the framework to two canonical quantum communication protocols that probe complementary aspects of performance: communication efficiency and communication security. To study efficiency, we analyze superdense coding with a fixed Bell-state alphabet under erasure, dephasing, and depolarizing channels. In all cases, the protocol capacity decreases monotonically with increasing atom-field coupling due to reduced coherence and distinguishability of the photonic codewords. In the strong-coupling limit, the four Bell states collapse into two distinct reduced photonic states, eliminating the second classical bit of ideal superdense coding. To study security, we incorporate the same intrinsic loss mechanism into the BB84 protocol. We introduce a modified protocol that discards erasure outcomes and use the microscopic model to estimate the intrinsic loss, providing a physical baseline for comparison with observed channel loss. We derive the asymptotic key rate and show that it decreases monotonically with increasing coupling strength, vanishing beyond a finite critical coupling. Taken together, these results show that interaction with a two-level atomic medium is a significant limiting factor for photonic quantum communication.

This paper is organized as follows. In Sec. \ref{sec:TheoryForSinglePhotoninteractingwithacollectionofatoms}, we develop the microscopic theory of a single photon interacting with an ensemble of two-level atoms and define the corresponding effective communication model. In Sec. \ref{section:SuperdenseCodingProtocol}, we study the effects of intrinsic and extrinsic errors on the efficiency of the superdense coding protocol. We review the ideal protocol and introduce the protocol capacity for the fixed Bell-state alphabet as a metric for information loss in \ref{subsection:ProtocolRecall}. In Sec. \ref{sec:EffectivePhotonicCommunicationModel}, we reformulate the superdense coding protocol in the language of our model and explain how intrinsic and extrinsic errors are incorporated. We model extrinsic errors as quantum channels, focusing specifically on erasure, depolarizing, and dephasing channels. In Sec. \ref{subsection:Capacity}, we explicitly calculate the capacity for each of these channels, accounting for atomic interactions. In Sec. \ref{section:QuantumKeyDistribution}, we investigate the impact of atomic interactions on the canonical BB84 quantum key distribution protocol and introduce the asymptotic key rate as a performance metric. In Sec. \ref{sec:protocolRecallBB84}, we revisit the BB84 protocol. In Sec. \ref{subsection:ProtocolModification}, we then modify the the  protocol to account for atomic interactions. In Sec. \ref{subsection:AsymptoticKeyRate}, we examine how atomic interactions affect the asymptotic key rate. Finally, in Sec. \ref{section:discussion}, we summarize our results and discuss future research directions.

\section{Theory of a Single Photon Interacting with a Collection of Atoms}
\label{sec:TheoryForSinglePhotoninteractingwithacollectionofatoms}
\subsection{Model}

We consider a model of the interaction between a quantized electromagnetic field and a collection of identical two-level atoms. We employ a real-space quantization framework for the electromagnetic field that explicitly incorporates polarization effects. Details of the quantization procedure is provided in Appendix \cref{Appendix:Real-spaceQuantizationofElectricMagneticField}. The total Hamiltonian consists of three terms:
\begin{equation}
H = H_F + H_A + H_I.
\end{equation}
The Hamiltonian of the field is given by 
\begin{equation}
\label{eq:photonicHamiltonian}
H_F = \hbar c\int d^3x\, (-\Delta)^{1/2} \boldsymbol{\phi}^\dagger(\bx)\cdot \boldsymbol{\phi}(\bx),
\end{equation}
where $c$ is the speed of light. The field operator $\boldsymbol{\phi}(\bx)$ is a three dimensional vector field satisfying $\nabla\cdot\boldsymbol{\phi}(\bx)=0$. The operators $\phi_j(\bx)$ and $\phi_j^\dag(\bx)$ annihilate and create a photon in a transverse spatial mode at $\bx$ with vector component $j$ respectively, and satisfy the commutation relations
\begin{equation}
[\phi_i(\bx),\phi_j^\dagger(\by)] = \delta^T_{ij}(\bx-\by),
\end{equation}
\begin{equation}
[\phi_i(\bx),\phi_j(\by)] = 0, \quad [\phi_i^\dagger(\bx),\phi_j^\dagger(\by)] = 0.
\end{equation} 
Here the transverse delta function $\delta_{ij}^T$ is defined by
\begin{equation}
\delta_{ij}^T(\bx) = \int \frac{d^3k}{(2\pi)^3} e^{i \bk \cdot \bx} \left(\delta_{ij} - \frac{k_i k_j}{\abs{\bk}^2}\right),
\end{equation}
and the  operator $(-\Delta)^{1/2}$ is defined by the Fourier integral
\begin{equation}
\begin{split}
\label{eq:DefinitionOfHalfLaplacian}
(-\Delta)^{1/2} f(\bx) &= \int \frac{d^3k}{(2\pi)^3} e^{i\bk \cdot \bx} \abs{\bk} \Tilde{f}(\bk),\\
\Tilde{f}(\bk)&=\int d^3 x e^{-i\bk \cdot \bx} f(\bx).
\end{split}
\end{equation}
The atomic Hamiltonian is given by
\begin{equation}
\label{eq:atomicHamiltonian}
H_A = \hbar \Omega \int d^3 x \ n(\bx) \hs^\dagger(\bx) \hs(\bx),
\end{equation}
where $n$ is the number density of atoms and $\Omega$ is the atomic transition frequency. The operators $\sigma(\bx)$ and $\sigma^\dag(\bx)$ annihilate and create an atomic excitation at position $\bx$ respectively, and obey the anticommutation relations
\begin{equation}
\{\sigma(\bx), \sigma^\dagger(\by)\} = \frac{\delta(\bx - \by)}{n(\bx)}, \quad  \{\sigma(\bx), \sigma(\by)\} = 0.
\end{equation}
We adopt anticommutation relations for convenience. In the single excitation sector considered here, the results are unchanged if the atomic excitations are instead described by bosonic field operators.
Within the rotating-wave and Markovian approximations, the atom-field interaction is described by the Hamiltonian
\begin{equation}
\begin{split}
\label{eq:interactionHamiltonian}
H_I = \hbar g \int d^3 x n(\bx) \hs^\dagger(\bx)  \hat{\boldsymbol{d}} \cdot \boldsymbol{\phi}(\bx) + {\rm h.c.},
\end{split}
\end{equation}
where $g$ is the coupling constant and $\hat{\boldsymbol{d}}$ is a  unit vector in the direction of the atomic dipole moment. In the Supplemental Material, we further investigate various aspects of this model.

The most general state of the system in the single-excitation sector can be written in the form
\begin{equation}
\ket{\Psi} = \int d^3x\, \qty[\psi(\bx)\cdot \boldsymbol{\phi}^\dagger(\bx) + n(\bx)a(\bx)\sigma^\dagger(\bx)]\ket{\emptyset},
\end{equation}
where $\ket{\emptyset} =\ket{0}_F \otimes \ket{0}_A$ denotes the combined vacuum state of the field and ground state of atoms. Here $\psi(\bx)$ is the probability amplitude for the field and $a(\bx)$ is the probability amplitude of exciting an atom at position $\bx$. The field amplitude is transverse, meaning that 
\begin{equation}
\label{eq:transverseCondition}
\nabla\cdot \psi(\bx)=0.
\end{equation}
Every state satisfies the normalization condition 
\begin{equation}
\label{eq:normalizationCondition}
\int d^3x\, \abs{\psi(\bx)}^2 + \int d^3x\, n(\bx)\abs{a(\bx)}^2 = 1 ,
\end{equation}
which ensures the conservation of probability

An energy eigenstate $\ket{\Psi}$ with energy $\hbar\omega$ satisfies the time-independent Schrödinger equation
\begin{equation}
H\ket{\Psi} = \hbar\omega\ket{\Psi}.
\end{equation}
It can then be seen that $\psi$ and $a$ obey the system of equations:
\begin{equation}
\begin{split}
\label{eq:FieldEquation}
&c(-\Delta)^{1/2}\psi(\bx) + g P_{T}(n a\hat{\boldsymbol{d}})(\bx)= \omega\psi(\bx),
\end{split}
\end{equation}
\begin{equation}
\begin{split}
\label{eq:atomFieldEquation}
&\Omega a(\bx) +g \psi(\bx) \cdot \hat{\boldsymbol{d}} = \omega a(\bx) .
\end{split}
\end{equation}
Here we have introduced the transverse projection operator $P_{T}$, which is defined for a vector field $F$ by
\begin{equation}
\label{eq:definitionOfPT}
(P_{T} F )_k(\bx) = \int d^3y \delta_{kl}^T(\bx - \by) F_l(\by).
\end{equation}

We now suppose the atomic number density is constant with $n(\bx) = n_0$. Eliminating the atomic amplitude $a$ from Eqs. \eqref{eq:FieldEquation} and \eqref{eq:atomFieldEquation}, we find that the one-photon amplitude $\psi$ solves the nonlinear eigenvalue problem
\begin{equation}
\begin{split}
\label{eq:nonlinearEigenvalueProblem}
&c(-\Delta)^{1/2} \psi(\bx) + \frac{n_0 {g}^2}{\omega-\Omega}P_{T}(\psi \cdot \hat{\boldsymbol{d}} \hat{\boldsymbol{d}})(\bx) = \omega \psi(\bx).\\
\end{split}
\end{equation}

It will prove useful to construct two solutions $\psi_1,\psi_2$ that satisfy  \cref{eq:nonlinearEigenvalueProblem} and obey \cref{eq:transverseCondition} and the orthogonality condition
\begin{equation}
\label{eq:orthogonalCondition}
\psi_1^\ast(\bx) \cdot \psi_2(\bx) = 0.
\end{equation}
Let $\hat{\bk}$ denote the direction of photon propagation and choose polarization vectors $\hat{\boldsymbol{\be}}_{\hat \bk,1}, \hat{\boldsymbol{\be}}_{\hat\bk,2}$ satisfying
\begin{equation}
\hat{\boldsymbol{\be}}_{\hat\bk s} \cdot \hat{\boldsymbol{ \be}}_{\hat\bk s'} = \delta_{ss'}, \quad \hat \bk \cdot \hat{\boldsymbol{\be}}_{\hat\bk s} = 0,
\end{equation}
for $s,s' = 1,2$. We consider the ansatz
\begin{equation}
\psi_j(\bx)
=\left(A_j\hat\be_{\hat{\bk},1}+B_j\hat\be_{\hat{\bk},2}\right)e^{i\bk_j\cdot\bx},
\end{equation}
where the wavevectors satisfy
\begin{equation}
\hat \bk_j=\hat \bk,
\end{equation}
for $j = 1,2$. Substituting this ansatz into \cref{eq:nonlinearEigenvalueProblem} yields the following condition on the coefficients:
\begin{equation}
\begin{split}
\label{eq:conditionOnSolutionCoefficient}
&(c\abs{\bk_j} - \omega)(A_j \hat{\be}_{\hat{\bk},1} + B_j \hat{\be}_{\hat{\bk},2})\\
&=-\frac{n_0g^2}{\omega - \Omega}(A_j \hat{\boldsymbol{d}}\cdot \hat{\be}_{\hat{\bk},1} + B_j \hat{\boldsymbol{d}}\cdot \hat{\be}_{\hat{\bk},2})(I - \hat{\bk}\otimes \hat{\bk}) \hat{\boldsymbol{d}},
\end{split}
\end{equation}
where $I - \hat{\bk}\otimes \hat{\bk}$ is the transverse projector.

We first consider the case in which $\hat{\boldsymbol{d}}\cdot \hat{\be}_{\hat{\bk},j} = 0$ for one value of $j$. Without loss of generality, we take $\hat{\boldsymbol{d}}\cdot \hat{\be}_{\hat{\bk},1} = 0$ since the polarization labels can always be interchanged. In this case, one can readily verify that $A_1 = 1, \ B_1 = 0$ and $A_2 = 0,  \ B_2 = 1$ provide the required coefficients. We next consider the case in which $\hat{\boldsymbol{d}}\cdot \hat{\be}_{\hat{\bk},j} \not= 0$ for both $j = 1, 2$. The first solution is then
\begin{equation}
A_1=\frac{C}{\hat{\boldsymbol{d}}\cdot\hat\be_{\hat{\bk},1}},
\quad
B_1=-\frac{C}{\hat{\boldsymbol{d}}\cdot\hat\be_{\hat{\bk},2}},
\end{equation}
where $C$ is the normalization constant chosen such that $\abs{A_1}^2 + \abs{B_1}^2 = 1$. Solving \cref{eq:conditionOnSolutionCoefficient} then determines the magnitude of the corresponding wavevector,
\begin{equation}
\abs{\bk_1}=\frac{\omega}{c}.
\end{equation}
Using the orthogonality condition \cref{eq:orthogonalCondition}, we obtain the second solution,
\begin{equation}
A_2=D\hat{\boldsymbol{d}}\cdot\hat\be_{\hat{\bk},1},
\quad
B_2=D\hat{\boldsymbol{d}}\cdot\hat\be_{\hat{\bk},2},
\end{equation}
where $D$ is a normalization constant fixed by $\abs{A_2}^2+\abs{B_2}^2 = 1$. Solving  \cref{eq:conditionOnSolutionCoefficient} determines the magnitude of the corresponding wavevector,
\begin{equation}
\abs{\bk_2} = \frac{\omega}{c} - \frac{n_0g^2}{c(\omega- \Omega)}\left[\left(\hat{\boldsymbol{d}} \cdot \hat{\be}_{\hat{\bk},1}\right)^2+ \left(\hat{\boldsymbol{d}} \cdot \hat{\be}_{\hat{\bk},2}\right)^2\right].
\end{equation}
For $j = 1,2$, we define
\begin{equation}
\hat{\boldsymbol{\epsilon}}_{\hat{\bk},j} = A_j\hat\be_{\hat{\bk},1}+B_j\hat\be_{\hat{\bk},2}.
\end{equation}
The two solutions can then be written compactly as
\begin{equation}
\label{eq:solution}
\psi_j(\bx) = \hat{\boldsymbol{\epsilon}}_{\hat{\bk},j} e^{i\bk_j\cdot \bx},
\end{equation}
for $j = 1,2$. When $g = 0$ or when the propagation direction is parallel to the dipole moment with $\hat{\boldsymbol{d}} = \hat{\bk}$, the two solutions are degenerate and $\bk_1 = \bk_2$. The atom-field interaction lifts this degeneracy by shifting the magnitude of the wave vector associated with the polarization component coupled to the atomic dipole.

\subsection{Communications Model}
\label{sec:EffectiveCommunicationModel}
In the preceding development, we formulated a microscopic theory describing the interaction of a single photon with a collection of two-level atoms. To apply this to the protocols of interest, we now translate the resulting photon-atom dynamics into a  communications model. We perform this task in two steps. First, we construct normalizable logical photonic qubit states from the plane-wave solutions of the microscopic model, encoding the logical information in polarization states. Second, we determine how the light-matter interaction transforms these logical states. In the protocol-specific applications considered in later sections, we trace out the atomic degrees of freedom for the relevant alphabet states. The resulting reduced photonic states describe the microscopic interaction as an intrinsic loss mechanism.

This reduction necessarily idealizes the measurement process. A fully realistic experimental treatment would require a more explicit specification of the measurement geometry and readout protocol. Depending on the platform, one may consider, for example, boundary measurements \cite{Fan2010,Roy2017} or integrated waveguide-based detection schemes \cite{Natarajan2013,Divochiy2008}. Recent work has also demonstrated waveguide-integrated, hybrid spatiotemporal-multiplexed SNSPD arrays, in which detector clicks are read out sequentially in time and the device provides access to both photon-number and position information \cite{Cheng2023}. In the present work, however, our goal is to isolate the loss generated solely by the light-matter interaction. We therefore assume ideal detection and neglect measurement inefficiency, finite timing resolution, and other readout effects. We further assume that the receiver knows the medium parameters and employs ideal detection matched to the $g$-dependent photonic mode. The resulting wave-vector shift is therefore not counted as a source of loss. 

The microscopic theory yields two plane-wave solutions for the photonic amplitude $\psi_j$, given in \cref{eq:solution}. Since physically meaningful states must be normalizable, whereas plane waves are not, we construct normalizable Gaussian wave packets centered on these plane-wave solutions,
\begin{equation}
\label{eq:normalizedWaveFunction}
\psi_j(\bx) = N_j\hat{\boldsymbol{\epsilon}}_{\hat{\bk},j}  \int_{\abs{\bk'} = \abs{\bk_j}} d\bk' e^{i \bk' \cdot \bx} e^{- \frac{\abs{\bk' - \bk_j}^2}{2\xi^2}},
\end{equation}
for $j = 1, 2.$ The parameter $\xi$ specifies the width of the Gaussian wave packet, and $N_j$ is a normalization constant chosen such that
\begin{equation}
\label{eq:normalizationConstant}
\int d^3x \psi_j^\ast(\bx)\cdot \psi_j(\bx) = 1.
\end{equation}

Using the normalized amplitude in \cref{eq:normalizedWaveFunction}, we define logical qubit states of the field by encoding the logical information in the polarization degrees of freedom:
\begin{equation}
\begin{split}
\label{eq:logicalStates}
\ket{0_L}_F &= N_1 \int d^3x  \int_{\abs{\bk'} = \abs{\bk_1}} d\bk' e^{i \bk' \cdot \bx} e^{- \frac{\abs{\bk' - \bk_1}^2}{2\xi^2}}\hat{\boldsymbol{\epsilon}}_{\hat{\bk},1} \cdot \boldsymbol{\phi}^\dagger(\bx)\ket{0}_F,\\
\ket{1_L}_F &= N_2 \int d^3x  \int_{\abs{\bk'} = \abs{\bk_2}} d\bk' e^{i \bk' \cdot \bx} e^{- \frac{\abs{\bk' - \bk_2}^2}{2\xi^2}}\hat{\boldsymbol{\epsilon}}_{\hat{\bk},2} \cdot \boldsymbol{\phi}^\dagger(\bx)\ket{0}_F.\\
\end{split}
\end{equation}
As shown in Appendix \cref{Appendix:OrthonormalConditionForLogicalStates}, these logical states form an orthonormal basis in the limit $\xi/\min\{\abs{\bk_1},\abs{\bk_2}\} \to 0$,
\begin{equation}
\bra{i_L}\ket{j_L}_F = \delta_{ij},
\end{equation}
for $i, j = 0, 1$.
For the subsequent analysis of loss channels, it is convenient to denote the vacuum state as an erasure or error state according to
\begin{equation}
\ket{e}_F = \ket{0}_F.
\end{equation}

We next determine how  light-matter interactions transforms the two logical states. By direct computation, we find that
 \begin{equation}
\hat{\boldsymbol{\epsilon}}_{\hat{\bk},1} \cdot \hat{\boldsymbol{d}} = 0,
\end{equation}
so $H_I$ acts trivially on $\ket{0_L}$, leaving this state unchanged. We therefore focus on the logical state $\ket{1_L}$, defined in \cref{eq:logicalStates}, before it enters the atomic medium. After the interaction, the state generally contains both a surviving one-photon component and an atomic-excitation component. We denote the complete interacting state in the single-excitation sector by $\ket{1_L(g)}$, which reduces to $\ket{1_L}$ when $g \to 0$.

To characterize the interacting state associated with $\ket{1_L(g)}$, we first specify its one-photon component. The logical encoding in \cref{eq:logicalStates} fixes the spatial wave-packet profile and polarization structure, while the coefficient $p(g)$ determines the weight of the surviving photonic component. The corresponding photonic amplitude is therefore
\begin{equation}
\label{eq:photonicAmplitudeWithInteraction}
\psi_2(\bx,g) = N_2\sqrt{p(g)}\hat{\boldsymbol{\epsilon}}_{\hat{\bk},2}  \int_{\abs{\bk'} = \abs{\bk_2}} d\bk' e^{i \bk' \cdot \bx} e^{- \frac{\abs{\bk' - \bk_2}^2}{2\xi^2}},
\end{equation}
where $p(g)$ is fixed by the normalization condition in \cref{eq:normalizationCondition}. Using the relation in \cref{eq:atomFieldEquation}, we find that
\begin{equation}
\begin{split}
p(g) &= \int d^3x \psi_2^\ast(\bx,g)\cdot \psi_2(\bx,g)\\
&= \frac{\abs{\omega - \Omega}^2}{\abs{\omega-\Omega}^2 + n_0 g^2 d_2^2},
\end{split}
\end{equation}
where we define $d_2 = \hat{\boldsymbol{d}} \cdot \hat{\boldsymbol{\epsilon}}_{\hat{\bk},2}$
Combining the above result with \cref{eq:photonicAmplitudeWithInteraction} and the atomic component determined by \cref{eq:atomFieldEquation}, the full interacting state can be written as 
\begin{equation}
\ket{1_L(g)} = \sqrt{p(g)}\ket{1_L}_F\otimes \ket{0}_A + \sqrt{1-p(g)} \ket{e}_F\otimes\ket{\bk_2}_A,
\end{equation}
where
\begin{equation}
\begin{split}
\ket{\bk_2}_A &= \sqrt{n_0}N_2\int d^3x \int_{\abs{\bk'} = \abs{\bk_2}} d\bk' e^{i\bk' \cdot \bx} e^{-\frac{\abs{\bk' - \bk_2}^2}{2\xi^2}}\hs^\dagger(\bx)\ket{0}_A.
\end{split}
\end{equation}
In the following sections, we apply this effective physical model to two specific communication protocols.

\section{Superdense Coding Protocol}
\label{section:SuperdenseCodingProtocol}
The superdense coding protocol provides a striking illustration of how quantum entanglement can enhance communication capacity. In the ideal setting, one party, named Alice, transmits two bits of classical information to a second party, Bob, by sending only a single qubit. The protocol begins with the distribution of an entangled pair of qubits, typically prepared in a Bell state, between Alice and Bob. Alice then encodes her two classical bits by applying one of four local unitary operations to her qubit. She subsequently sends that qubit to Bob, who performs a joint Bell measurement on the pair to determine unambiguously which operation Alice applied, thereby decoding the two bits of information. In practice, the performance of this protocol is limited by the quality of the quantum channel and the atom-field interaction, and its effectiveness can be quantified rigorously by the protocol capacity for the Bell alphabet. In this section, we first review the ideal protocol in the absence of atom-field coupling. We then incorporate the interaction, construct the corresponding communication model, and formulate the resulting encoding task. Finally, we evaluate the capacity and quantify the effect of the light-matter interaction.

\subsection{Protocol Recall}
\label{subsection:ProtocolRecall}
We first describe the protocol in the ideal, lossless case, which employs the four Bell states
\begin{equation}
\begin{split}
\label{eq:definitionofBellStates}
\ket{\Psi_{00}} &= \frac{\ket{0_L}_A\otimes \ket{0_L}_B + \ket{1_L}_A\otimes \ket{1_L}_B}{\sqrt{2}},\\
\ket{\Psi_{01}} &= \frac{\ket{0_L}_A\otimes \ket{1_L}_B + \ket{1_L}_A\otimes \ket{0_L}_B}{\sqrt{2}},\\
\ket{\Psi_{10}} &= \frac{\ket{0_L}_A\otimes \ket{0_L}_B - \ket{1_L}_A\otimes \ket{1_L}_B}{\sqrt{2}},\\
\ket{\Psi_{11}} &= \frac{\ket{0_L}_A\otimes \ket{1_L}_B - \ket{1_L}_A\otimes \ket{0_L}_B}{\sqrt{2}}.\\
\end{split}
\end{equation}
We consider an initial state in which Alice and Bob share an entangled photon pair. Although the protocol can be implemented using any Bell state, we assume for concreteness that the shared initial state is $\ket{\Psi_{00}}$. To encode two classical bits, Alice applies one of four local unitary operations to her qubit:
\begin{equation}
\begin{split}
&00 \longrightarrow Id_A \otimes Id_B \ket{\Psi_{00}} = \ket{\Psi_{00}}\\
&01 \longrightarrow X_A \otimes Id_B \ket{\Psi_{00}} = \ket{\Psi_{01}}\\
&10 \longrightarrow Z_A \otimes Id_B \ket{\Psi_{00}} = \ket{\Psi_{10}}\\
&11 \longrightarrow Z_AX_A \otimes Id_B \ket{\Psi_{00}} = \ket{\Psi_{11}},\\
\end{split}
\end{equation}
where $Id_A$ and $Id_B$ are the identity operators on Alice's and Bob's photons, respectively, and $X_A$ and $Z_A$ are the Pauli operators acting on Alice's logical qubit. If $\sigma_{AB}$ is the density matrix of the two-qubit system after Alice's operation, she then sends her photon to Bob through a quantum channel, which we model with a completely positive trace-preserving (CPTP) map $\mathcal{Q}_A$. The final state available to Bob is $\mathcal{Q}_A\otimes Id_B(\sigma_{AB})$, as his photon does not pass through the channel. 

As a baseline calculation, we first consider a perfect channel, $\mathcal{Q}_A = Id_A$, which represents lossless communication. The classical capacity of a quantum channel $\mathcal{Q}$ is given by the Holevo-Schumacher-Westmoreland (HSW) theorem as the maximum of the Holevo information $\chi$ over all possible input state ensembles  $\{(k_i, \sigma_i)\}$:
\begin{equation}
\begin{split}
C(\mathcal{Q})&= \max_{\{(k_i,\sigma_i)\}} \chi(\mathcal{Q},\{(k_i,\sigma_i)\})\\
&= \max_{\{(k_i,\sigma_i)\}} \left[S(\mathcal{Q}(\sum_i k_i \sigma_i)) - \sum_i k_i S(\mathcal{Q}(\sigma_i)) \right],
\end{split}
\end{equation}
where $\sum_i k_i =1$ and $S$ is the von Neumann entropy. Since the entropy of a pure state is zero, we directly have
\begin{equation}
C(Id_A\otimes Id_B) = 2.
\end{equation}
This recovers the well-known result for ideal superdense coding: the transmission of a single qubit enables the communication of two classical bits.

\subsection{Communication Model}
\label{sec:EffectivePhotonicCommunicationModel}
In the physical setting we consider, the superdense coding protocol employs two photons, with information encoded in their polarization degrees of freedom. During transmission, Alice's photon interacts with the atomic medium, whereas Bob's photon remains isolated in an ideal environment. In the preceding section, we derived an effective communication model from the microscopic Hamiltonian, describing how the logical states of Alice's photon are transformed by the atom-field interaction. We now use these results to determine how the Bell-state alphabet is modified by the light-matter interaction and then trace over the unobserved atomic degrees of freedom. The resulting reduced photonic states represent the microscopic interaction as an intrinsic loss mechanism. We subsequently combine this intrinsic loss with standard external noise channels acting on Alice's transmitted photon, thereby obtaining the effective channel models used to calculate the superdense-coding capacity.

From \cref{sec:EffectiveCommunicationModel}, the logical states of Alice's photon are
\begin{equation}
\ket{0_L(g)}_A = \ket{0_L}_A\otimes \ket{0}_A,
\end{equation}
\begin{equation}
\ket{1_L(g)}_A = \sqrt{p(g)}\ket{1_L}_A\otimes \ket{0}_A + \sqrt{1-p(g)} \ket{e}_A\otimes\ket{\bk_2}_A.
\end{equation}
Since Bob's photon remains in an ideal environment and is independent of Alice's atom-field system, the four interacting Bell states can be constructed directly as
\begin{equation}
\begin{split}
\ket{\Psi_{00}(g)} &= \frac{\ket{0_L(g)}_A\otimes\ket{0_L}_B + \ket{1_L(g)}_A \otimes \ket{1_L}_B}{\sqrt{2}},\\
\ket{\Psi_{01}(g)} &= \frac{\ket{0_L(g)}_A\otimes \ket{1_L}_B + \ket{1_L(g)}_A\otimes \ket{0_L}_B}{\sqrt{2}},\\
\ket{\Psi_{10}(g)} &= \frac{\ket{0_L(g)}_A\otimes \ket{0_L}_B - \ket{1_L(g)}_A\otimes \ket{1_L}_B}{\sqrt{2}},\\
\ket{\Psi_{11}(g)} &= \frac{\ket{0_L(g)}_A\otimes \ket{1_L}_B - \ket{1_L(g)}_A\otimes \ket{0_L}_B}{\sqrt{2}}.\\
\end{split}
\end{equation}

The full photon-atom system is described by the pure-state density matrix $\ket{\Psi_{ij}(g)}\bra{\Psi_{ij}(g)}$, where $i,j=0,1$. Since only the photonic degrees of freedom are  accessible, we trace over the atomic state space. This procedure yields the following reduced photonic density matrices:
\begin{equation}
\begin{split}
\sigma^{00}(g) &= \Tr_A(\ket{\Psi_{00} (g)}\bra{\Psi_{00} (g)} )\\
&=\frac{1}{2}(\ket{0_L}_A\otimes \ket{0_L}_B + \sqrt{p(g)}\ket{1_L}_A\otimes \ket{1_L}_B)\\
&(\bra{0_L}_A\otimes \bra{0_L}_B + \sqrt{p(g)}\bra{1_L}_A\otimes \bra{1_L}_B)\\
&+\frac{1-p(g)}{2}\ket{e}\bra{e}_A\otimes \ket{1_L}\bra{1_L}_B,
\end{split}
\end{equation}
\begin{equation}
\begin{split}
\sigma^{01}(g) &= \Tr_A(\ket{\Psi_{01} (g)}\bra{\Psi_{01} (g)} )\\
&=\frac{1}{2}(\ket{0_L}_A\otimes \ket{1_L}_B + \sqrt{p(g)}\ket{1_L}_A\otimes \ket{0_L}_B)\\
&(\bra{0_L}_A\otimes \bra{1_L}_B + \sqrt{p(g)}\bra{1_L}_A\otimes \bra{0_L}_B)\\
&+\frac{1-p(g)}{2}\ket{e}\bra{e}_A\otimes \ket{0_L}\bra{0_L}_B,
\end{split}
\end{equation}
\begin{equation}
\begin{split}
\sigma^{10}(g) &= \Tr_A(\ket{\Psi_{10} (g)}\bra{\Psi_{10} (g)} )\\
&=\frac{1}{2}(\ket{0_L}_A\otimes \ket{0_L}_B - \sqrt{p(g)}\ket{1_L}_A\otimes \ket{1_L}_B)\\
&(\bra{0_L}_A\otimes \bra{0_L}_B - \sqrt{p(g)}\bra{1_L}_A\otimes \bra{1_L}_B)\\
&+\frac{1-p(g)}{2}\ket{e}\bra{e}_A\otimes \ket{1_L}\bra{1_L}_B,
\end{split}
\end{equation}
\begin{equation}
\begin{split}
\sigma^{11}(g) &= \Tr_A(\ket{\Psi_{11} (g)}\bra{\Psi_{11} (g)} )\\
&=\frac{1}{2}(\ket{0_L}_A\otimes \ket{1_L}_B - \sqrt{p(g)}\ket{1_L}_A\otimes \ket{0_L}_B)\\
&(\bra{0_L}_A\otimes \bra{1_L}_B - \sqrt{p(g)}\bra{1_L}_A\otimes \bra{0_L}_B)\\
&+\frac{1-p(g)}{2}\ket{e}\bra{e}_A\otimes \ket{0_L}\bra{0_L}_B.
\end{split}
\end{equation}
The reduced density matrices incorporate the loss induced by the microscopic light-matter interaction. To complete the communication model, we now include external transmission noise acting on Alice's photon. These external losses are represented by completely positive trace-preserving (CPTP) maps, while Bob's photon is left unchanged. A dominant loss mechanism in photonic communication is photon erasure, where a photon is irrecoverably lost from the channel. We model this process using an erasure channel $\mathcal{E}_A$ which acts on Alice's qubit according to
\begin{equation}
\mathcal{E}_A(\sigma_A) = s \sigma_A + (1-s)\ket{e}\bra{e}_A,
\end{equation}
where $s \in [0,1]$ is the transmission rate and $\ket{e}_A$ is the erasure (vacuum) state. As detailed in the Appendix \cref{appendix:theoremForFixedDensityMatrixQuantumChannel}, applying this channel to Alice's system while leaving Bob's system intact yields the following density matrix
\begin{equation}
\mathcal{E}_A \otimes Id_B(\sigma_{AB}) = s \sigma_{AB} + (1-s) \ket{e}\bra{e}_A \otimes \Tr_A(\sigma_{AB}). 
\end{equation}

We also analyze the impact of a single-qubit dephasing channel $\mathcal{D}_A$ which introduces phase errors with probability $\epsilon \in [0,1]$: 
\begin{equation}
\mathcal{D}_A(\sigma_A) = (1-\epsilon) \sigma_A + \epsilon Z_A \sigma_A Z_A.
\end{equation}
Applying this map to Alice's photon while Bob's remains unaffected results in the final state: 
\begin{equation}
\begin{split}
\mathcal{D}_A \otimes Id_B(\sigma_{AB}) &= (1-\epsilon) \sigma_{AB} + \epsilon Z_A\otimes Id_B \sigma_{AB} Z_A \otimes Id_B.
\end{split}
\end{equation}

Last, we consider the effect of a depolarizing channel, $\mathcal{M}_A$. This channel models a scenario where, with probability $\epsilon$, the environment replaces the original state with a maximally mixed state: 
\begin{equation}
\label{eq:DefinitionOfDepolarizingChannel}
\mathcal{M}_A(\sigma_A) = (1-\epsilon) \sigma_A + \epsilon \frac{Id_A}{3},
\end{equation}
where $Id_A = \ket{0_L}\bra{0_L}_A+\ket{1_L}\bra{1_L}_A+\ket{e}\bra{e}_A.$
Applying the result established in Appendix \cref{appendix:theoremForFixedDensityMatrixQuantumChannel}, the action of this channel on the joint state is
\begin{equation}
\begin{split}
\mathcal{M}_A \otimes Id_B(\sigma_{AB}) &= (1-\epsilon)\sigma_{AB} + \epsilon \frac{Id_A}{3}\otimes \Tr_A(\sigma_{AB}),
\end{split}
\end{equation}

With these models for both intrinsic and external decoherence in place, the primary objective of the next section is to calculate the superdense-coding capacity for the fixed Bell-state alphabet under the combined effects of the atomic interaction and these different quantum channels.

\subsection{Capacity}
\label{subsection:Capacity}
We now calculate the superdense-coding capacity for the fixed Bell-state alphabet by solving the optimization problem defined by the HSW theorem. The optimization is not performed over arbitrary photon input states or over arbitrary encodings beyond the standard four dense-coding messages. Instead, the four input states associated with Alice's possible messages are fixed:
\begin{equation}
\sigma^{00}(g), \ \sigma^{01}(g), \ \sigma^{10}(g), \ \sigma^{11}(g). 
\end{equation}
We therefore consider a general ensemble $\{(k_{ij}, \sigma^{ij}(g))\}_{i,j = 0,1}$ subject to the constraint $\sum_{i,j = 0,1} k_{ij} = 1$, and seek  the distribution ${k_{ij}}$ that maximizes the Holevo information
$$\chi(\mathcal{Q}_A \otimes Id_B,\{(k_{ij}, \sigma^{ij}(g))\}_{i,j = 0,1})$$
for a fixed coupling strength $g$.

For the erasure channel, the capacity of the superdense coding protocol with the fixed Bell-state alphabet in the presence of the atomic interaction is
\begin{equation}
\begin{split}
C(\mathcal{E}_A \otimes Id_B,g) 
&=-\frac{s}{2}\log_2\left(\frac{s}{4}\right)
 -\frac{sp(g)}{2}\log_2\left(\frac{sp(g)}{4}\right)\\
&-\frac{2-s(1+p(g))}{2}\log_2\frac{2-s(1+p(g))}{4}\\
&+\frac{s(1+p(g))}{2}\log_2\frac{s(1+p(g))}{2}\\
&+\frac{1-s}{2}\log_2\frac{1-s}{2}\\
&+\frac{1-sp(g)}{2}\log_2\frac{1-sp(g)}{2}.
\end{split}
\end{equation}
The details of the calculations are given in the Appendix \cref{Appendix:CapacityForErasureChannel}.

\begin{figure}[t]
\centering
\includegraphics[width=1 \linewidth]{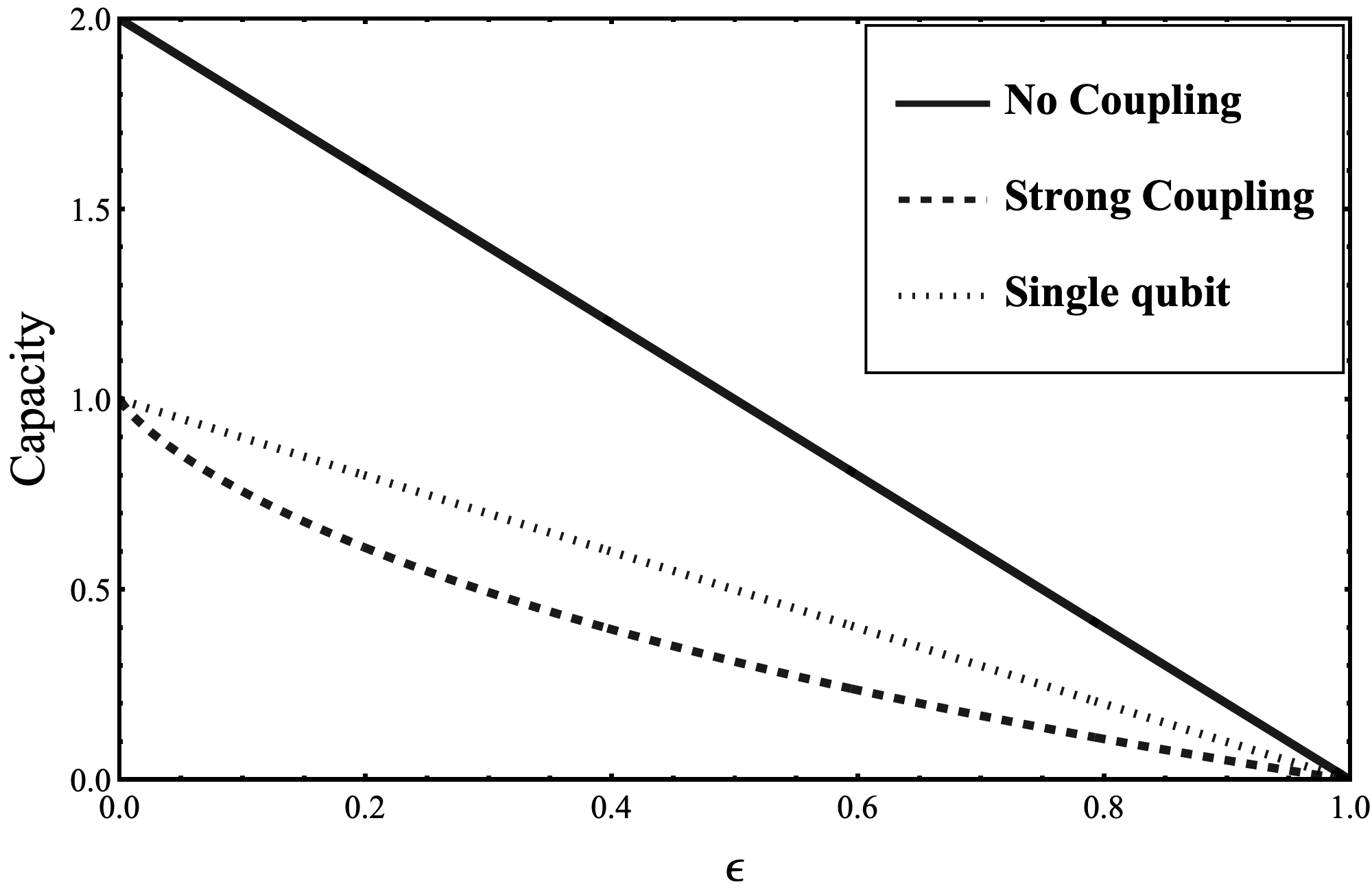}
\caption{Erasure channel protocol capacity for the Bell alphabet.\label{fig:ErasureChannelCapacity}}
\end{figure} 

Analyzing the limiting cases of the above result provides further insight. First, in the absence of any light-matter coupling ($g=0$), the probability $ p(g)$ becomes 1 and the capacity simplifies to $C = 2s$. If we define the erasure probability as $\epsilon = 1-s$, the capacity is $2(1-\epsilon)$. This recovers the known result for the entanglement-assisted classical capacity of an erasure channel \cite{Bennett99,Hsieh10,Wilde_2013}, which was derived without reference to a specific physical model. Our approach therefore correctly reproduces the canonical result in the appropriate limit.
Second, for complete erasure ($s=0$), the capacity vanishes independent of the coupling strength. This behavior is expected since no photon reaches the receiver.
Finally, we consider the worst-case strong-coupling limit ($g \to \infty$), in which the  capacity becomes
\begin{equation}
\begin{split}
&\lim_{g\to \infty}C(\mathcal{E}_A\otimes Id_B,g)\\
&= 1-\frac{2-s}{2}\log_{2}(2-s)
+\frac{1-s}{2}\log_{2}(1-s).
\end{split}
\end{equation}
The no-coupling and strong-coupling limits are compared in \cref{fig:ErasureChannelCapacity} with the single-qubit classical capacity as a function of the error rate $\epsilon$. The results show that, in the strong-coupling regime, the entanglement-assisted advantage of superdense coding is completely lost. Moreover, the resulting capacity is lower than that obtained by directly transmitting classical information with a single photon. In particular, for $0<\epsilon<1$, the strong-coupling erasure capacity lies strictly below the single-qubit erasure capacity. The origin of this reduction becomes clear by examining the codewords in the strong-coupling limit. As $g \to \infty$, the surviving photonic probability vanishes:
\begin{equation}
\lim_{g\to \infty} p(g) = 0.
\end{equation}
The four-state Bell alphabet thus
collapses into two distinct states,
\begin{equation}
\begin{split}
\label{eq:strongCouplingStates}
\rho_0
&=\frac{1}{2}\ket{00}\bra{00}
 +\frac{1}{2}\ket{e1}\bra{e1},\\
\rho_1
&=\frac{1}{2}\ket{01}\bra{01}
 +\frac{1}{2}\ket{e0}\bra{e0}.
\end{split}
\end{equation}
The phase-related encodings within each pair are  indistinguishable, eliminating the second classical bit supplied by ideal superdense coding. At $\epsilon=0$, the two remaining states have orthogonal supports and can still encode one bit perfectly. For nonzero $\epsilon$, however, both the medium-induced and channel-induced erasure processes populate the same flag sector $|e\rangle$. The resulting output states therefore have overlapping supports, making the two surviving messages only partially distinguishable. This overlap explains why the fixed-alphabet superdense-coding capacity falls below $1-\epsilon$.

The dephasing channel exhibits related behavior. The capacity of the superdense-coding protocol for the fixed Bell-state inputs, including the effect of the atomic interaction, is
\begin{equation}
\begin{split}
&C(\mathcal{D}_A \otimes Id_B,g) \\
&= \frac{3+p(g)-p(g)\log_2(p(g))}{2}\\
&+\lambda_{+}\log_{2}\lambda_{+}+\lambda_{-}\log_{2}\lambda_{-},
\end{split}
\end{equation}
where
\begin{equation}
\lambda_\pm = \frac{\frac{1+p}{2}\pm \sqrt{(\frac{1-p}{2})^2+(1-2\epsilon)^2p}}{2}.
\end{equation}
The  derivation of this result can be found in Appendix \cref{Appendix:CapacityForDephasingChannel}.
\begin{figure}[t]
\centering
\includegraphics[width=1 \linewidth]{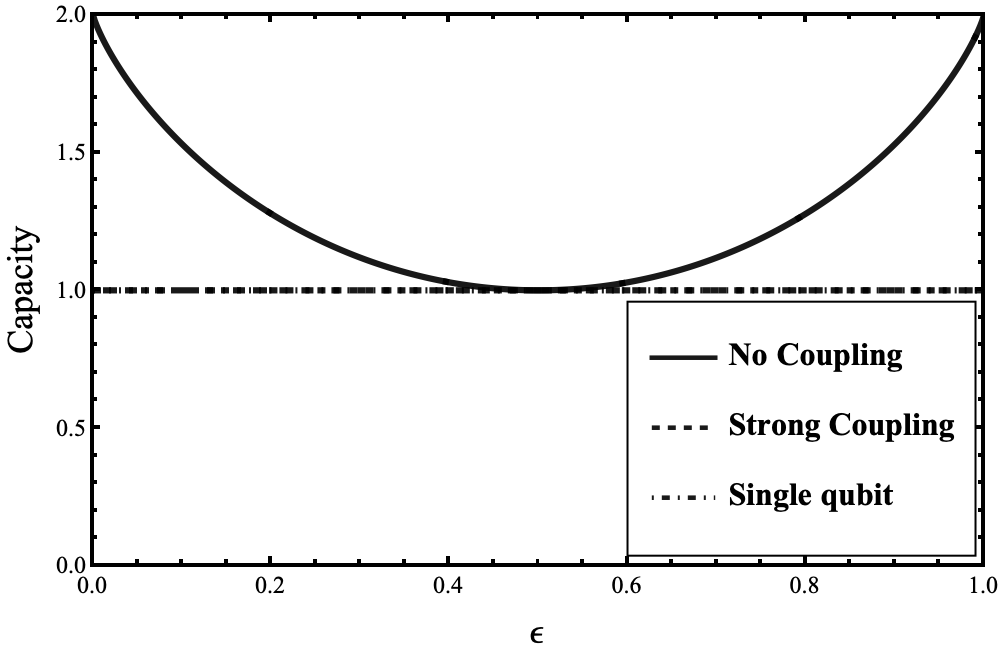}
\caption{Dephasing channel protocol capacity for the Bell alphabet.\label{fig:DephasingChannelCapacity}}
\end{figure} 

As in the erasure-channel case, the no-coupling limit ($g=0$) correctly reproduces the known entanglement-assisted classical capacity of a single-qubit dephasing channel \cite{Wilde_2013}.

A more intriguing behavior emerges when we compare the strong coupling capacity to the classical capacity of a single qubit sent through the same dephasing channel. For a single qubit, the capacity is well-known to be 1 \cite{Wilde_2013}. This is achieved by encoding information exclusively in the computational basis ($\ket{0}, \ket{1}$), which is immune to phase-flip errors. In the strong-coupling limit, as $g \to \infty$, and the Bell alphabet collapses into the two states given in \cref{eq:strongCouplingStates}. These states are independent of the dephasing error for every value of $\epsilon$. Consequently, the strong-coupling protocol capacity coincides with the single-qubit classical capacity, as shown in \cref{fig:DephasingChannelCapacity}.

For the depolarizing channel, the capacity of the superdense-coding protocol with the fixed Bell-state alphabet with atomic interactions is
\begin{equation}
\begin{split}
&C(\mathcal{M}_A\otimes Id_B,g) \\
&= -\left(\frac{1-\epsilon}{2}+\frac{\epsilon}{3}\right)\log_2\left(\frac{1-\epsilon}{4}+\frac{\epsilon}{6}\right)\\
&-\left(\frac{1-\epsilon}{2}p(g)+\frac{\epsilon}{3}\right)\log_2\left(\frac{1-\epsilon}{4}p(g)+\frac{\epsilon}{6}\right)\\
&-\left(\frac{1-\epsilon}{2}(1-p(g))+\frac{\epsilon}{3}\right)\log_2\left(\frac{1-\epsilon}{4}(1-p(g))+\frac{\epsilon}{6}\right)\\
&+\frac{2\epsilon}{3}\log_2(\frac{\epsilon}{6})\\
&+\left[\frac{(1-\epsilon)(1-p(g))}{2}+\frac{\epsilon}{6}\right]\log_2\left[\frac{(1-\epsilon)(1-p(g))}{2}+\frac{\epsilon}{6}\right]\\
&+\left[\frac{(1-\epsilon)(1+p(g))}{2}+\frac{\epsilon}{6}\right]\log_2\left[\frac{(1-\epsilon)(1+p(g))}{2}+\frac{\epsilon}{6}\right].
\end{split}
\end{equation}
The  derivation is provided in Appendix \cref{Appendix:CapacityForDepolarizingChannel}.
In the absence of coupling ($g = 0$), the above expression does not reduce to the known entanglement-assisted classical capacity of a conventional single-qubit depolarizing channel. This difference arises since the maximally mixed state used in the definition of the depolarizing channel in \cref{eq:DefinitionOfDepolarizingChannel} includes the erasure state. 

\begin{figure}[t]
\centering
\includegraphics[width=1 \linewidth]{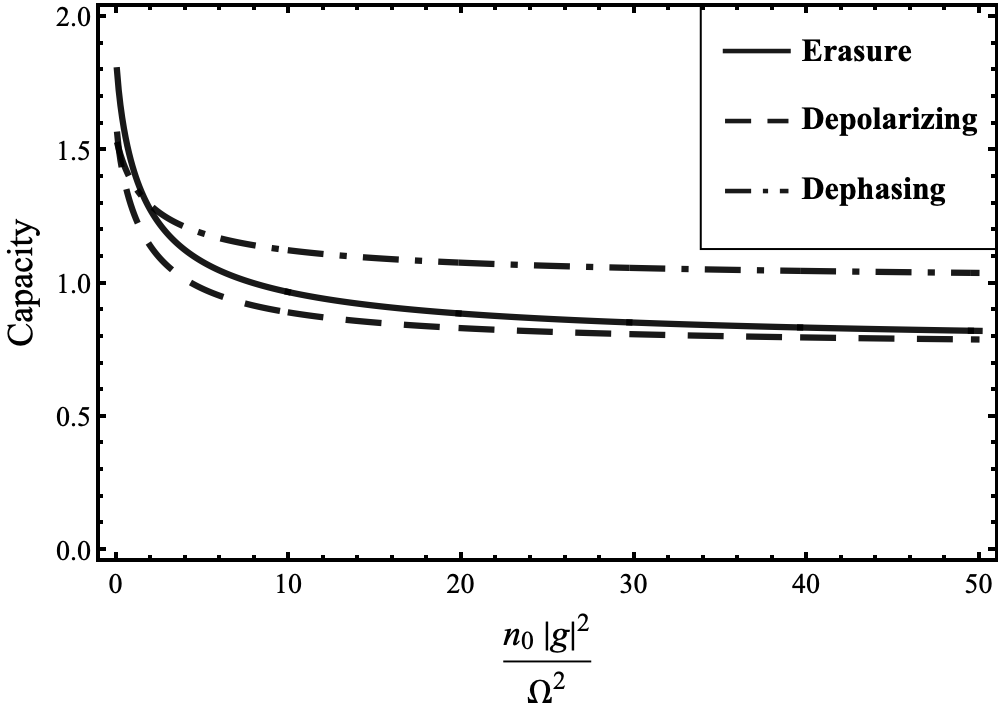}
\caption{Channel protocol capacities for the Bell alphabet as a function of coupling strength.\label{fig:CapacityDrops}}
\end{figure} 

To consolidate our findings, we plot the capacities for all three channels as functions of the dimensionless coupling strength ${n_0g^2}/{\Omega^2}$ at a fixed error rate $\epsilon$. We choose the dimensionless parameters $\epsilon = 0.2, \ {\Omega}/{\omega} =2, \ d_2 = 0.5$. As shown in \cref{fig:CapacityDrops}, the capacities of all three channels decrease monotonically as the coupling strength increases. This behavior demonstrates that the atom-field interaction has a detrimental effect on quantum-information transmission. More specifically, increasing the coupling transfers a larger fraction of the excitation from the accessible photonic subsystem to the unobserved atomic subsystem, thereby reducing the distinguishability of the output codewords and the amount of classical information recoverable by the receiver.

\section{Quantum Key Distribution BB84 Protocol}
\label{section:QuantumKeyDistribution}
A central challenge in cryptography is the secure distribution of secret keys between distant parties. Quantum key distribution (QKD) addresses this problem quantum mechanically. In the BB84 protocol, for example, information is encoded in two non-orthogonal bases, so that an eavesdropper who measures the transmitted states generally introduces disturbances that can be detected by legitimate users.

In any practical implementation, however, photon loss is unavoidable. From the users' perspective, a missing detection event does not by itself reveal whether the signal was lost through ordinary propagation or removed through adversarial intervention. For this reason, standard security analyses of BB84 are typically formulated conservatively: uncharacterized loss is treated in a worst-case manner, and the final key rate is bounded accordingly.

A microscopic model of information transmission  cannot replace such worst-case security analysis, but it can provide useful calibration information. In the present work, the relevant physical process is the interaction between the photon and an atomic medium. Within this model, we can estimate the loss rate directly from the underlying light-matter dynamics and thereby obtain a prediction for the loss expected in the absence of adversarial intervention.
The role of this estimate does not by itself establish security against arbitrary attacks, nor does it rule out attacks that reproduce the expected loss while acting on the surviving signals. Rather, it provides a physical baseline for evaluating whether the observed behavior is consistent with the assumed channel model. If the observed loss significantly exceeds this baseline, then the channel behavior may be treated as grounds for aborting the protocol or for adopting more conservative security parameters.

Accordingly, the analysis in this section should be understood within a restricted channel model. Our goal is not to derive an unconditional security proof for BB84 in the presence of a general adversary. Instead, we show that a physical model of photon loss can be incorporated into the protocol description. Under the assumptions stated in the Appendix \cref{Appendix:RestrictedSecurity}, we also provide a restricted security result for arbitrary quantum attacks on the surviving photonic systems.

It can be argued that nondetection events are harmless since the corresponding bits are simply discarded. This reasoning, however, implicitly assumes that loss is independent of the encoded bit and basis choices, and that the random numbers used for state preparation and measurement are private and sufficiently uncorrelated \cite{Gottesman2004,Maroy2010,Bouda2012,Li2015}. However, these assumptions can fail in practical implementations: imperfect randomness may give an observer Eve partial information about the encoding choices, while loss or postselection can be exploited through efficiency-mismatch or time-shift-type attacks \cite{Makarov2006,Zhao2008}. A sufficiently sophisticated eavesdropper could then selectively suppress or retain signals in a way correlated with the underlying bit values, so that even partial information about the discarded patterns may compromise the final key. For this reason, it is essential to minimize the rate of unexplained loss and to account for all relevant physical loss mechanisms.

\subsection{Protocol Recall}
\label{sec:protocolRecallBB84}
We first describe the BB84 protocol in the ideal setting (without atom-field interactions). The protocol uses a single photon as a qubit which encodes the polarization degree of freedom and requires the following four states:
\begin{equation}
\ket{0_L}_F, \ket{1_L}_F, \ket{\pm_L}_F =\frac{\ket{0_L}_F\pm \ket{1_L}_F}{\sqrt{2}}.
\end{equation}

We now describe the standard BB84 protocol for a generic quantum channel $\mathcal{Q}$.

\textit{BB84 Protocol:} (1) Alice generates $(4+\delta_1)n$ random bits. (2) Alice generates a random $(4+\delta_1)n$-bit string $b$. For each bit, she prepares $\{\ket{0_L}_F, \ket{1_L}_F\}$ if the corresponding bit of $b$ is $0$ or $\{\ket{+_L}_F, \ket{-_L}_F\}$ if the corresponding bit of $b$ is $1$. The collection of qubits is denoted as $\{\rho_i\}_{i = 0}^{(4+\delta_1)n}$ (3) Alice sends the resulting qubits $\{\rho_i\}_{i = 0}^{(4+\delta_1)n}$ to Bob through a quantum channel $\mathcal{Q}$ individually. (4) Bob receives the $(4+\delta_1)n$ qubits $\{\mathcal{Q}(\rho_i)\}_{i = 0}^{(4+\delta_1)n}$ measuring each qubit in the $X$ or $Z$ basis at random. (5) Alice announces $b$. (6) Alice and Bob discard any bits where Bob measured a different basis than Alice prepared. With high probability, there are at least $2n$ bits left (if not, abort the protocol). (7) Alice randomly selects a subset of $n$ bits that will serve as a check on Eve's interference, and publicly tells Bob which bits she selects. (8) Alice and Bob publicly announce and compare the values of their $n$ check bits. If they are more than the threshold, which depends on the characteristics of the channel $\mathcal{Q}$, disagree, they abort the protocol. (9) Alice and Bob perform information reconciliation and privacy amplification on the remaining $n$ bits to obtain $m$ shared key bits.

Two details of this protocol warrant clarification. First, the parameter $\delta_1$ ensures that a sufficient number of bits survive the sifting process, where Alice and Bob compare their measurement choices. By choosing a large enough $\delta_1$, the probability of retaining at least $2n$ bits can be made arbitrarily close to one. In the asymptotic limit of large $n$, $\delta_1$ can be chosen to approach zero. Second, the decision to abort in step (8) depends on a pre-agreed error threshold. This threshold is not arbitrary but is determined by the physical characteristics of the quantum channel $\mathcal{Q}$. Specifically, it is calculated from the expected fidelities between the pure states sent by Alice and the mixed states received by Bob. For an input state $\ket{i}$, the corresponding fidelity is 
\begin{equation} F_i = F(\ket{i}\bra{i}, \mathcal{Q}(\ket{i}\bra{i}) ), \end{equation} for $i \in {0,1,+,-}$, where the fidelity between a pure state $\ket{\Psi}$ and a mixed state $\rho$ is given by \begin{equation} F(\ket{\Psi}\bra{\Psi}, \rho) = \bra{\Psi}\rho \ket{\Psi}. 
\end{equation}
The performance of the protocol is measured by the asymptotic key rate $R$, defined as the number of final secret key bits generated per initial signal transmitted, in the limit of a large number of signals:
\begin{equation}
R = \lim_{l \to \infty} \frac{k(l)}{l},
\end{equation} 
where $l$ is the total number of signals sent by Alice and $k(l)$ is the length of the final secret key. For the BB84 protocol, a well-known lower bound on the key rate \cite{Shor00} is given by 
\begin{equation}
R = \max\{0,p_sp_p(1 - 2 H(\epsilon))\},
\end{equation}
where $H(x)$ is the binary entropy function
\begin{equation}
H(x) = - x \log_2(x) - (1-x) \log_2(1-x).
\end{equation}
This result is related to the quantum Gilbert-Varshamov bound for Calderbank-Shor-Steane (CSS) codes. The term $p_s$ is the probability that a signal is retained after the basis-sifting step, while $p_p$ is the probability that a sifted bit is part of the final key rather than the check sequence. In the protocol described, this corresponds to $p_s = 1/2$ and $p_p = 1/2$. The quantity $\epsilon$ is the quantum bit error rate (QBER) of the channel, defined as 
\begin{equation}
\epsilon = \max\{ 1-\frac{F_0 + F_1}{2}, 1-\frac{F_+ + F_-}{2}\},
\end{equation}
representing the maximum error rate between the bit-flip and phase-flip channels. 
We will see that, when the atom-field interaction is included, this key rate is modified.

\subsection{Protocol Modification}
\label{subsection:ProtocolModification}
From \cref{sec:EffectiveCommunicationModel}, the computational-basis logical states with an atom-field interaction of strength $g$ are
\begin{equation}
\ket{0_L(g)} = \ket{0_L}_F\otimes \ket{0}_A,
\end{equation}
\begin{equation}
\ket{1_L(g)} = \sqrt{p(g)}\ket{1_L}_F\otimes \ket{0}_A + \sqrt{1-p(g)} \ket{e}_F\otimes\ket{\bk_2}_A.
\end{equation}
Accordingly, we define the Hadamard-basis states  as
\begin{equation}
\ket{+_L(g)} = \frac{\ket{0_L(g)} + \ket{1_L(g)}}{\sqrt{2}},
\end{equation}
and
\begin{equation}
\ket{-_L(g)} = \frac{\ket{0_L(g)} - \ket{1_L(g)}}{\sqrt{2}}.
\end{equation}
Since Bob's measurement  probes only the photon field, the effective state available to him is obtained by tracing over the unobserved atomic degrees of freedom. The resulting mixed state is given by
\begin{equation}
\begin{split}
\sigma^i(g) &= \Tr_A(\ket{i_L(g)}\bra{i_L(g)}),\\
\end{split}
\end{equation}
for $i = 0,1,+,-$. More explicitly, these states are
\begin{equation}
\sigma^0(g) = \ket{0_L}\bra{0_L},
\end{equation}
\begin{equation}
\sigma^1(g) = p(g)\ket{1_L}\bra{1_L} + (1-p(g))\ket{e}\bra{e},
\end{equation}
\begin{equation}
\begin{split}
\sigma^+(g) &= \frac{1}{2}(\ket{0_L} + \sqrt{p(g)} \ket{1_L})(\bra{0_L} + \sqrt{p(g)} \bra{1_L})\\
&+\frac{1-p(g)}{2}\ket{e}\bra{e},
\end{split}
\end{equation}
\begin{equation}
\begin{split}
\sigma^-(g) &= \frac{1}{2}(\ket{0_L} - \sqrt{p(g)} \ket{1_L})(\bra{0_L} - \sqrt{p(g)}\bra{1_L})\\
&+\frac{1-p(g)}{2}\ket{e}\bra{e}.
\end{split}
\end{equation}

We are now ready to introduce the modified protocol.

\textit{Modified BB84 Protocol:} (1) Alice generates $(4+\delta_1 + \delta_2)n$ random bits. (2) Alice generates a random $(4+\delta_1+ \delta_2)n$-bit string $b$. For each bit, she prepares $\{\ket{0_L}_F, \ket{1_L}_F\}$ if the corresponding bit of $b$ is $0$ or $\{\ket{+_L}_F, \ket{-_L}_F\}$ if the corresponding bit of $b$ is $1$. The collection of qubits is denoted  $\{\rho_i\}_{i = 0}^{(4+\delta_1+ \delta_2)n}$ (3) Alice sends the resulting qubits $\{\rho_i\}_{i = 0}^{(4+\delta_1+ \delta_2)n}$ to Bob through a quantum channel $\mathcal{Q}$ with coupling strength $g$. (4) Bob receives the $(4+\delta_1+ \delta_2)n$ qubits $\{\mathcal{Q}(\sigma^i(g))\}_{i = 0}^{(4+\delta_1+ \delta_2)n}$, measuring each qubit in the $X$ or $Z$ basis at random. (5) Alice announces $b$. (6) Alice and Bob discard any bits where Bob measured a different basis than Alice prepared. Moreover, Alice and Bob discard any result where Bob measures an erasure state. With high probability, there are at least $2n$ bits left (if not, abort the protocol). (7) Alice randomly selects a subset of $n$ bits that will serve as a check on Eve's interference, and tells Bob publicly which bits she selects. (8) Alice and Bob publicly announce and compare the values of their $n$ check bits. If more than a fixed threshold amount, which depends on the characteristics of the channel $\mathcal{Q}$, disagree, they abort the protocol. (9) Alice and Bob perform information reconciliation and privacy amplification on the remaining $n$ bits to obtain $m$ shared key bits.

In this modified protocol, $\delta_1$ serves the same purpose as before, accounting for the overhead of basis sifting. The new parameter, $\delta_2$, is introduced to compensate for the expected fraction of signals lost to erasure events. For the protocol to succeed with high probability, $\delta_2$ must be chosen to be slightly larger than the value
\begin{equation}
\frac{4}{1-p_e(g)} - 4-\delta_1,
\end{equation}
where $p_e(g)$ is the probability that Bob receives an erasure state. In the limit $n \to \infty$, $\delta_2$ approaches this value $\frac{4}{1-p_e(g)} - 4$. Assuming Alice chooses her encoding states with uniform probability, the erasure probability is the average over all four input states:
\begin{equation}
\label{eq:definitionOfErasureProbability}
p_e(g) = \frac{1}{4} \sum_{i \in \{0,1, +, -\} } \bra{e} \mathcal{Q}(\sigma^i(g)) \ket{e}.
\end{equation}

The crucial modification to the above protocol, compared to the ideal case, is the post-selection in step (6), where Alice and Bob discard all outcomes corresponding to a non-detection event. We now calculate the expected intrinsic loss rate of the channel and set the parameter $\delta_2$ accordingly. Any significant deviation from this benchmark can then be attributed to the actions of an eavesdropper. In contrast, the standard BB84 protocol does not account for the physical origin of loss. An eavesdropper could potentially mask their intercept-resend activity within the channel's overall loss rate.

\subsection{Asymptotic Key Rate}
\label{subsection:AsymptoticKeyRate}
In this section, we derive the asymptotic key rate of the modified BB84 protocol as a function of the coupling strength $g$. In Appendix \cref{Appendix:RestrictedSecurity}, we provide the corresponding restricted security claim and a sketch of its proof. The security result is restricted since Eve is assumed to have no access to the atomic degrees of freedom. The additional assumptions required for the proof are also stated in that appendix. Under these assumptions, the asymptotic key rate is
\begin{equation}
R(g) = \max\{0,p_s(g)p_p(1 - 2 H(\epsilon(g)))\}.
\end{equation}

The primary change to the key rate formula arises from the post-selection against erasure events. The effective sifting probability, $p_s(g)$, is now the product of the basis-sifting probability (1/2) and the probability that the transmitted qubit is not erased. This gives 
\begin{equation}
\label{eq:modifiedSiftingProbability}
p_s(g) = \frac{1}{2}(1-p_e(g)),
\end{equation}
where $p_e(g)$ is the total erasure probability defined on \cref{eq:definitionOfErasureProbability}. The fraction of the key used for parameter estimation $p_p$ remains $1/2$. The other significant change is to the quantum bit error rate (QBER) $\epsilon$. Since all erasure events are discarded, the QBER must be calculated based on the conditional probability of an error, given that a conclusive result was obtained. This is given by 
\begin{equation}
\begin{split}
\label{eq:modifiedErrorRate}
\epsilon = \max\left\{1-\frac{F^{(0)}_\mathcal{Q}(g) + F^{(1)}_\mathcal{Q}(g)}{2-E^{(0)}_\mathcal{Q}(g)-E^{(1)}_\mathcal{Q}(g)},\right. \\
\left. 1- \frac{F^{(+)}_\mathcal{Q}(g) + F^{(-)}_\mathcal{Q}(g)}{2-E^{(+)}_\mathcal{Q}(g)-E^{(-)}_\mathcal{Q}(g)}\right\},
\end{split}
\end{equation}
where we define the fidelity for each of the input states for $i \in \{0,1, + , -\}$,
\begin{equation}
F^{(i)}_\mathcal{Q}(g) = \bra{i_L}\mathcal{Q}(\sigma^{i}(g)) \ket{i_L},
\end{equation}
and the corresponding erasure probability as
\begin{equation}
E^{(i)}_\mathcal{Q}(g) = \bra{e}\mathcal{Q}(\sigma^{i}(g)) \ket{e}.
\end{equation}

To obtain a more explicit result, we model the external channel as an identity channel. This choice isolates the effect of the atom-field interaction on the asymptotic key rate.
We begin by calculating the total erasure probability generated by the atom-field interaction. Direct evaluation of \cref{eq:definitionOfErasureProbability} gives
\begin{equation}
p_e(g) = \frac{1-p(g)}{2}.
\end{equation}
The effective sifting probability is then
\begin{equation}
p_s = \frac{1+p(g)}{4}.
\end{equation}
The conditional error rate is
\begin{equation}
\epsilon = \frac{(1-\sqrt{p(g)})^2}{2(1+p(g))}.
\end{equation}
Substituting these quantities into the key-rate expression yields
\begin{equation}
\label{eq:asymptoticKeyRate}
R(g)  = \max\{0, \frac{1+p(g)}{8}(1-2H(\frac{(1-\sqrt{p(g)})^2}{2(1+p(g))}))\}.
\end{equation}

\begin{figure}[t]
\centering
\includegraphics[width=1 \linewidth]{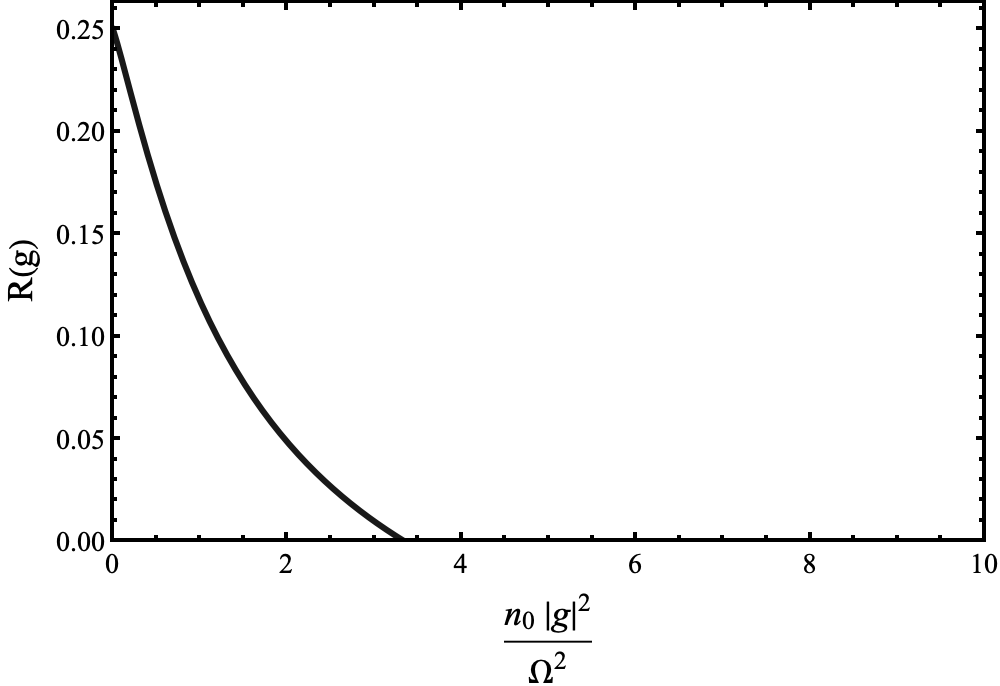}
\caption{Coupling-dependence of asymptotic key rate.\label{fig:AsymptoticKeyRate}}
\end{figure}

To illustrate this result, we plot the key rate as a function of the dimensionless coupling strength $n_0g^2/\Omega^2$, while fixing the remaining dimensionless parameters to be $\Omega/\omega = 2, d_2 = 0.5$. As shown in \cref{fig:AsymptoticKeyRate}, the key rate decreases monotonically with increasing coupling strength. In the absence of atom-field interactions ($g=0$), we have $ p(0) =1$, and the key rate becomes $ R(0) = 1/4$, recovering the standard result for an identity channel. The key rate vanishes at the critical coupling strength
\begin{equation}
\frac{n_0g^2}{\Omega^2} = \frac{\abs{\omega-\Omega}^2}{\Omega^2d^2_2} \frac{4\sqrt{\epsilon_\ast(1-\epsilon_\ast)}}{1-2\sqrt{\epsilon_\ast(1-\epsilon_\ast)}} ,
\end{equation}
where $\epsilon_\ast$ is the root for $H(\epsilon_\ast) = \frac{1}{2}$ and $0 < \epsilon_\ast < \frac{1}{2}$ and approximately $\epsilon_\ast = 0.11$.
For coupling strengths greater than or equal to this critical value, the asymptotic key rate is zero. Thus, sufficiently strong atom-field coupling prevents secret-key generation within the present channel model. Increasing the coupling both reduces the fraction of conclusive photonic detections and increases the conditional error rate among the surviving signals, ultimately eliminating the positive key rate.

\section{Discussion}
\label{section:discussion}
In this work, we have developed a physical framework for analyzing photonic quantum communication protocols in an interacting atomic medium. By modeling a quantized electromagnetic field coupled to a collection of two-level atoms, our formalism provides a unified approach for incorporating both intrinsic errors arising from the light-matter interaction and extrinsic errors associated with external loss channels. We have applied this framework to investigate both the efficiency of quantum communication, through the superdense coding protocol, and its security, through the BB84 quantum key distribution protocol. Although the model is idealized and does not include a detailed treatment of state preparation or measurement, it isolates the limitations imposed by photon-atom interactions and therefore provides a benchmark for photonic communication through atomic media.

For superdense coding, we evaluated the capacity of the protocol for the fixed Bell-state alphabet under erasure, dephasing, and depolarizing channels, finding that in all three cases the capacity decreases monotonically with increasing atom-field coupling. This reduction originates from the decreased survival probability $p(g)$ of the polarization mode coupled to the atomic dipole. That is, as the coupling increases, an increasingly large fraction of the excitation is transferred to the atomic subsystem, reducing both the coherence of the photonic codewords and their distinguishability at the receiver. In the absence of coupling ($g=0$), the results for the erasure and dephasing channels reduce to  the corresponding standard entanglement-assisted capacities. The strong-coupling limit further clarifies how the atomic interaction eliminates the dense-coding advantage. As $g \to \infty$, the survival probability approaches zero, and the four Bell-state codewords collapse into two distinct reduced photonic states, rendering the two encodings within each phase-related pair indistinguishable and eliminating the second classical bit available in ideal superdense coding. The effect of additional channel noise then depends on its structure: for the erasure channel, the strong-coupling capacity lies below the direct single-qubit erasure capacity for $0<\epsilon<1$, since intrinsic and external erasure populate the same flag sector, making the two surviving messages ambiguous; for the dephasing channel, the two remaining codewords survive phase errors, and the strong-coupling capacity therefore coincides with the single-qubit classical capacity. These results demonstrate that the loss of the entanglement-assisted advantage is determined not only by the strength of the microscopic interaction but also by the manner in which the resulting reduced codewords are affected by the external noise process.

For BB84, we incorporated the same microscopic loss mechanism into a modified protocol that discards erasure outcomes before parameter estimation and key generation. The model provides an a priori estimate of the intrinsic loss expected from the photon-atom interaction and thus supplies a physical baseline against which the observed loss can be compared. A statistically significant excess over this prediction may indicate that the assumed channel model is incomplete and may justify aborting the protocol or adopting more conservative security parameters. Within this framework, the asymptotic key rate decreases monotonically with increasing coupling strength since the atom-field interaction both lowers the probability of obtaining a conclusive photonic detection and increases the conditional error rate among the surviving signals. In the absence of coupling, the result reduces to the standard value $R(0) = 1/4$ for the protocol convention adopted here. At a finite critical coupling strength, the conditional QBER reaches the threshold beyond which the lower bound on the asymptotic key rate vanishes. Thus, sufficiently strong light-matter coupling does not merely reduce the number of retained signals; it ultimately prevents secret-key generation within the present model.

Several promising directions for future research are now apparent. The present formalism can be applied directly to other communication schemes, such as the entanglement-based E91 QKD protocol, and can be adapted to protocols that encode information in different degrees of freedom, including photon-number, spatial, and time-bin encoding. In addition, the real-space quantization framework can be extended naturally to more complex and experimentally relevant geometries, such as waveguides, enabling detailed studies of mode-dependent loss and crosstalk. The same approach could also be generalized to arrays of coupled waveguides, providing a useful platform for investigating the challenges associated with scaling quantum communication networks. From a more practical perspective, the formalism can be extended to incorporate fluctuations in atomic density. Another important direction is the inclusion of time dependent states to capture dynamical phenomena such as Rabi oscillations. Such an extension would likely require going beyond the Markovian approximation in order to include memory effects and other features that become important in dense atomic ensembles. Similarly, replacing the uniform dipole orientation assumed here with a statistical distribution of dipole moments would allow the study of orientational disorder. Ensemble averaging, potentially combined with a mean-field approximation, could then reveal how microscopic disorder modifies polarization-dependent loss and communication performance. Finally, an important next step is to investigate how these physically motivated error models can inform the design of more efficient and robust quantum error-correction codes tailored to the distinct noise signatures generated by fundamental light-matter interactions.

\begin{acknowledgments}
We thank Steven M. Girvin and Imran M. Mirza for helpful discussions. John Schotland was partially supported by the Simons Foundation grant SFI-MPS-TSRG-00023182-04. Generative AI (Claude and ChatGPT) was used to assist with the development and verification of mathematical arguments, drafting and revision of the text, and preparation
of the figures. It was also used for code development and numerical computations. The authors take full responsibility for the results of this study and the final form of the manuscript.
\end{acknowledgments}

\appendix
\section{Real-space Quantization}
\label{Appendix:Real-spaceQuantizationofElectricMagneticField}
We begin with the standard electromagnetic-field Hamiltonian
\begin{equation}
\label{eq:FieldHamiltonian}
H = \int d^3k \sum_s \hbar \omega_\bk \, a^\dagger_{\bk s}a_{\bk s},
\end{equation}
where $\omega_\bk = c\abs{\bk}$ and $a_{\bk s}$ ($a^\dag_{\bk s}$) annihilates (creates) a photon of momentum $\bk$ and transverse polarization $s$. In addition, the following commutation relations are satisfied:
\begin{equation}
\label{eq:commutationRelation}
[a_{\bk s},a^\dagger_{\bk' s'}] = \delta(\bk-\bk')\delta_{ss'},
\end{equation}
\begin{equation}
[a_{\bk s},a_{\bk' s'}] = 0,
\end{equation}
for $s \in \{1,2\}$.
For each nonzero $\bk$, we choose polarization vectors $\hat{\boldsymbol{\epsilon}}_{\bk,1}, \hat{\boldsymbol{\epsilon}}_{\bk,2}$, where
\begin{equation}
\hat{\boldsymbol{\epsilon}}_{\hat\bk s} \cdot \hat{\boldsymbol{ \epsilon}}_{\hat\bk s'} = \delta_{ss'}, \quad \hat \bk \cdot \hat{\boldsymbol{\epsilon}}_{\hat\bk s} = 0,
\end{equation}
for $s,s' \in \{1,2\}$. 

To carry out the quantization of the field in real-space, we define the operator
\begin{equation}
\boldsymbol{\hp}(\bx) = \int \frac{d^3k}{(2\pi)^{3/2}} e^{i\bk\cdot\bx} \sum_s\hat{\boldsymbol{\epsilon}}_{\hat \bk, s}a_{\bk s}.
\end{equation}
Making use of \cref{eq:commutationRelation}, we find $\boldsymbol{\hp}(\bx)$ obeys the commutation relations
\begin{equation}
\begin{split}
[\hp_{i}(\bx), \hp^\dagger_{j} (\bx')] &= \delta_{ij}^T(\bx - \bx') ,\\ [\hp_{i}(\bx), \hp_{i} (\bx')] &= 0,
\end{split}
\end{equation}
where the transverse projection operator is defined by
\begin{equation}
\delta_{ij}^T(\bx) = \int \frac{d^3k}{(2\pi)^3} e^{-i \bk \cdot \bx} \left(\delta_{ij} - \frac{k_i k_j}{\abs{\bk}^2}\right).
\end{equation} 
We note that $\nabla\cdot \boldsymbol{\hp}(\bx) = 0$ and thus
\begin{equation}
P_T(\boldsymbol{\hp}) =\boldsymbol{\hp}.
\end{equation}
Utilizing the above results, we find that \cref{eq:FieldHamiltonian} becomes
\begin{equation}
H = \hbar c \int d^3x\, (-\Delta)^{1/2}\boldsymbol{\hp}^\dagger(\bx)\cdot \boldsymbol{\hp}(\bx),
\end{equation}
where $(-\Delta)^{1/2}$ is defined by \cref{eq:DefinitionOfHalfLaplacian}.

\section{Logical States}
\label{Appendix:OrthonormalConditionForLogicalStates}
We  consider the logical states
\begin{equation}
\begin{split}
&\ket{i_L}_F\\
&= N_{i+1} \int d^3x \int_{\abs{\bp} = \abs{\bk_{i+1}}} d\bp e^{i \bp \cdot \bx} e^{- \frac{\abs{\bp - \bk_{i+1}}^2}{2\xi^2}}  \hat{\boldsymbol{\epsilon}}_{\hat{\bk},i+1}\cdot \hp^\dagger(\bx)\ket{0}_F,
\end{split}
\end{equation}
for $i = 0,1$.
We will show that in the limit $\xi \to 0$,
\begin{equation}
\bra{j_L}\ket{i_L}_F = \delta_{ij} .
\end{equation}
This follows from the following calculation:
\begin{equation}
\begin{split}
&\bra{j_L}\ket{i_L}_F\\
&=N_{i+1}N_{j+1}(2\pi)^3\int d^3p \delta(\abs{\bp}-\abs{\bk_{i+1}}) \delta(\abs{\bp}-\abs{\bk_{j+1}})\\ &\times e^{-\frac{\abs{\bp - \bk_{i+1}}^2+ \abs{\bp - \bk_{j+1}}^2}{2\xi^2}} \hat{\epsilon}_{\hat{\bk},i+1} \cdot(I - \hat{\bp}\otimes \hat{\bp})\hat{\epsilon}_{\hat{\bk},j+1}.
\end{split}
\end{equation}
If $\abs{\bk_{i+1}} \not= \abs{\bk_{j+1}}$, we immediately have $\bra{j_L}\ket{i_L}_F = 0.$
If $\abs{\bk_{i+1}} = \abs{\bk_{j+1}}= \lambda$ then
\begin{equation}
\begin{split}
&\bra{j_L}\ket{i_L}_F\\
&= N_{i+1}N_{j+1} (2\pi)^3\int d^3p e^{-\frac{\abs{\bp -\bk}^2}{\xi^2}} \delta(\abs{\bp} - \lambda)\delta(\abs{\bp} - \lambda) \\
&\times\hat{\boldsymbol{\epsilon}}_{\hat{\bk}, j+1} \cdot (I- \hat \bp \otimes \hat \bp)\hat{\boldsymbol{\epsilon}}_{\hat{\bk}, i+1}.\\
\end{split}
\end{equation}
Here we regularize the divergence by setting $\delta(0) = {1}/{2\pi \xi}$. We find that
\begin{equation}
\begin{split}
\bra{j_L}\ket{i_L}_F &= \frac{2 e^{-\lambda^2/\xi^2} \xi^4 \delta_{ij}}{\left(1 - e^{-4\lambda^2/\xi^2}\right)\lambda^4}\\
&\times \left(-\frac{\lambda^2 \cosh\left(\lambda^2/\xi^2\right)}{\xi^2} + \left(1 + \frac{\lambda^4}{\xi^4}\right)\sinh\left(\lambda^2/\xi^2\right)\right).
\end{split}
\end{equation}
In the limit ${\xi} \to 0$ we have
\begin{equation}
\bra{j_L}\ket{i_L}_F = \delta_{ij}.
\end{equation}

\section{A Quantum Channel Identity}
\label{appendix:theoremForFixedDensityMatrixQuantumChannel}
We consider a quantum channel $Q$ of the form
\begin{equation*}
 Q[\rho]=p\rho+(1-p)\sigma\Tr(\rho),
\end{equation*}
for a fixed density matrix $\sigma$. Then, for a two qubit system $A\otimes B$,
\begin{equation*}
Q_A\otimes\Id_B[\rho_{AB}]
 =p\rho_{AB}+(1-p)\sigma\otimes\Tr_A(\rho_{AB}).
\end{equation*}

The proof is as follows. Let $\rho_{AB}$ be a mixed state on $A\otimes B$. Then we can write
\begin{equation*}
 \rho_{AB}=\sum_{i,j}\ket{i}\bra{j}_A\otimes X_{ij},
\end{equation*}
where $\{\ket{i}_A\}$ are basis of $A$ and $X_{ij}$ are some operator on $B$. Then,
\begin{equation*}
 \Tr_A(\rho_{AB})=\sum_i X_{ii}.
\end{equation*}
Now, we compute explicitly $Q$ acting on $\ket{i}\bra{j}$:
\begin{equation*}
 Q[\ket{i}\bra{j}]=p\ket{i}\bra{j}+(1-p)\sigma\,\delta_{ij}.
\end{equation*}
Then
\begin{align*}
Q_A\otimes\Id_B[\rho_{AB}]
&=\sum_{i,j}Q_A[\ket{i}\bra{j}_A]\otimes\Id_B X_{ij}\\
&=\sum_{i,j}\left[p\ket{i}\bra{j}_A+(1-p)\sigma\delta_{ij}\right]\otimes X_{ij}\\
&=p\rho_{AB}+(1-p)\sigma\otimes\sum_{i,j}\delta_{ij}X_{ij}\\
&=p\rho_{AB}+(1-p)\sigma\otimes\sum_iX_{ii}\\
&=p\rho_{AB}+(1-p)\sigma\otimes\Tr_A(\rho_{AB}).
\end{align*}

\section{Capacity for Erasure Channel}
\label{Appendix:CapacityForErasureChannel}
Here we calculate the capacity of the erasure channel with superdense coding accounting for atomic interactions. The first step is to calculate the entropy of the output states. Applying the combined intrinsic and external loss channel to a single input state yields 
\begin{equation}
\begin{split}
&\mathcal{E}_{A}\otimes \Id_{B}(\sigma^{00}(g))\\
&=\frac{s}{2}\left(\ket{0_L}_A\otimes\ket{0_L}_B+\sqrt{p(g)}\ket{1_L}_A\otimes\ket{1_L}_B\right)\\
&\left(\bra{0_L}_A\otimes\bra{0_L}_B
+\sqrt{p(g)}\bra{1_L}_A\otimes\bra{1_L}_B\right)\\
&\quad+\frac{1-s}{2}\ket{e}_A\bra{e}\otimes \left(\ket{0_L}_B\bra{0_L}+p(g)\ket{1_L}_B\bra{1_L}\right)\\
&\quad+\frac{1-p(g)}{2}\ket{e}_A\bra{e}\otimes\ket{1_L}_B\bra{1_L} ,
\end{split}
\end{equation}
which has eigenvalues 
$\frac{s(1+p)}{2}, \frac{1-s}{2},\frac{1-sp}{2}.$
This gives the entropy
\begin{equation}
\begin{split}
&S(\mathcal{E}_A \otimes Id_B(\sigma^{ij}(g)) )\\
&=-\frac{s(1+p)}{2}\log_2\frac{s(1+p)}{2}
 -\frac{1-s}{2}\log_2\frac{1-s}{2}\\
&\quad-\frac{1-sp}{2}\log_2\frac{1-sp}{2}.
\end{split}
\end{equation}
Next, we consider the average input state which is $\rho_{AB}(g) = \sum_{i,j = 0,1} k_{ij} \sigma^{ij}(g)$, and maximize the Holevo information for the ensemble, which is achieved when $k_{ij} = \frac{1}{4}$. This gives the capacity of the erasure channel, accounting for the influence of the atomic medium:
\begin{equation}
\begin{split}
&C(\mathcal{E}_A\otimes Id_B,g)\\
&=-\frac{s}{2}\log_2\left(\frac{s}{4}\right)
 -\frac{sp}{2}\log_2\left(\frac{sp}{4}\right)\\
&\quad-\frac{2-s(1+p)}{2}\log_2\frac{2-s(1+p)}{4}\\
&\quad+\frac{s(1+p)}{2}\log_2\frac{s(1+p)}{2}\\
&\quad+\frac{1-s}{2}\log_2\frac{1-s}{2}
 +\frac{1-sp}{2}\log_2\frac{1-sp}{2}.
\end{split}
\end{equation}

\section{Capacity For Dephasing Channel}
\label{Appendix:CapacityForDephasingChannel}
Here we calculate the capacity of the dephasing channel used in conjunction with superdense coding accounting for atomic interactions. First, we calculate the entropy of the individual output states. Applying the channel map to a single input state $\sigma^{00}(g)$ yields
\begin{equation}
\begin{split}
\mathcal{D}_{A}\otimes\Id_{B}(\sigma^{00}(g))
&=\frac{1}{2}\ket{00}\bra{00}+\frac{p}{2}\ket{11}\bra{11}\\
&+\frac{(1-2\epsilon)\sqrt p}{2}\ket{11}\bra{00}\\
&+\frac{(1-2\epsilon)\sqrt p}{2}\ket{00}\bra{11}\\
&+\frac{1-p}{2}\ket{e}_A\bra{e}\otimes\ket{1_L}_B\bra{1_L},\\
\end{split}
\end{equation}
with eigenvalues $\frac{1-p}{2},\lambda_{\pm}$, where
\begin{equation}
\lambda_\pm = \frac{\frac{1+p}{2}\pm \sqrt{(\frac{1-p}{2})^2+(1-2\epsilon)^2p}}{2}.
\end{equation}
The Bell states input have the same eigenvalues and thus the entropy is given by
\begin{equation}
\begin{split}
&S(\mathcal{D}_A \otimes Id_B(\sigma^{ij}(g)) )\\
&=-\frac{1-p}{2}\log_2\left(\frac{1-p}{2}\right)\\
&-\lambda_{+}\log_{2}\lambda_{+}-\lambda_{-}\log_{2}\lambda_{-}.
\end{split}
\end{equation}
Next, we calculate the maximum of the Holevo information of the average input state $\rho_{AB}(g) = \sum_{i,j = 0,1} k_{ij} \sigma^{ij}(g)$:
\begin{equation}
\begin{split}
&C(\mathcal{D}_A\otimes Id_B,g)\\
&= \frac{3+p(g)-p(g)\log_2(p(g))}{2}\\
&+\lambda_{+}\log_{2}\lambda_{+}+\lambda_{-}\log_{2}\lambda_{-},
\end{split}
\end{equation}
which is obtained when the input probabilities are uniform, i.e., $k_{ij} = 1/4$ for $i,j = 0,1$. 

\section{Capacity For Depolarizing Channel}
\label{Appendix:CapacityForDepolarizingChannel}
Here we calculate the capacity for the depolarizing channel with superdense coding. First, we obtain the entropy of the individual output states. Applying the channel map to a single input state $\sigma^{00}(g)$ yields
\begin{equation}
\begin{split}
&\M_{A}\otimes\Id_{B}(\sigma^{00}(g))\\
&=\frac{1-\epsilon}{2}(\ket{00}+\sqrt p\ket{11})(\bra{00}+\sqrt p\bra{11})\\
&+\frac{(1-\epsilon)(1-p)}{2}\ket{e}\bra{e}\otimes\ket{1}\bra{1}\\
&\quad+\frac{\epsilon}{6}
\left[\ket{0}\bra{0}+\ket{1}\bra{1}+\ket{e}\bra{e}\right]\otimes\left[\ket{0}\bra{0}+\ket{1}\bra{1}\right],\\
\end{split}
\end{equation} 
which has eigenvalues 
\begin{equation}
\frac{\epsilon}{6}, \ \frac{\epsilon}{6}, \ \frac{\epsilon}{6}, \ \frac{\epsilon}{6}, \ \frac{(1-\epsilon)(1-p)}{2}+\frac{\epsilon}{6}, 
\frac{(1-\epsilon)(1+p)}{2}+\frac{\epsilon}{6} .
\end{equation}
All four Bell states give rise to the same eigenvalues. The corresponding von Neumann entropy is
\begin{equation}
\begin{split}
&S(\mathcal{M}_A \otimes Id_B(\sigma^{ij}(g)) )\\
&= -\frac{2\epsilon}{3}\log_2(\frac{\epsilon}{6})\\
&-\left[\frac{(1-\epsilon)(1-p)}{2}+\frac{\epsilon}{6}\right]\log_2\left[\frac{(1-\epsilon)(1-p)}{2}+\frac{\epsilon}{6}\right]\\
&-\left[\frac{(1-\epsilon)(1+p)}{2}+\frac{\epsilon}{6}\right]\log_2\left[\frac{(1-\epsilon)(1+p)}{2}+\frac{\epsilon}{6}\right],
\end{split}
\end{equation}
for $i,j = 0,1$.
Next, we calculate the maximum Holevo information of the average output state $\rho_{AB}(g) = \sum_{i,j = 0,1} k_{ij} \sigma^{ij}(g)$, which is given by
\begin{equation}
\begin{split}
&C(\mathcal{M}_A\otimes Id_B,g) \\
&= -\left(\frac{1-\epsilon}{2}+\frac{\epsilon}{3}\right)\log_2\left(\frac{1-\epsilon}{4}+\frac{\epsilon}{6}\right)\\
&-\left(\frac{1-\epsilon}{2}p(g)+\frac{\epsilon}{3}\right)\log_2\left(\frac{1-\epsilon}{4}p(g)+\frac{\epsilon}{6}\right)\\
&-\left(\frac{1-\epsilon}{2}(1-p(g))+\frac{\epsilon}{3}\right)\log_2\left(\frac{1-\epsilon}{4}(1-p(g))+\frac{\epsilon}{6}\right)\\
&+\frac{2\epsilon}{3}\log_2(\frac{\epsilon}{6})\\
&+\left[\frac{(1-\epsilon)(1-p(g))}{2}+\frac{\epsilon}{6}\right]\log_2\left[\frac{(1-\epsilon)(1-p(g))}{2}+\frac{\epsilon}{6}\right]\\
&+\left[\frac{(1-\epsilon)(1+p(g))}{2}+\frac{\epsilon}{6}\right]\log_2\left[\frac{(1-\epsilon)(1+p(g))}{2}+\frac{\epsilon}{6}\right] .
\end{split}
\end{equation}

\section{Restricted Security of BB84 and Asymptotic Key Rate}
\label{Appendix:RestrictedSecurity}

We denote the logical single-photon qubit space by
$\mathcal{H}_L=\operatorname{span}\{\ket{0_L},\ket{1_L}\}$
and let $\ket{e}$ be an erasure state satisfying $\ket{e}\perp\mathcal{H}_L$.

We make the following assumptions:
\begin{enumerate}
\item The source emits exactly one ideal logical photon.

\item The microscopic atom--field interaction is trusted, fixed, and memoryless, and it occurs before Eve can act. Eve cannot control the atoms or obtain information from the atomic degrees of freedom.

\item Alice and Bob have complete knowledge of the microscopic description of the system, including the exact atom--field coupling.

\item All random numbers used in the protocol are private, independent, and uniformly distributed. The classical communication channel is authenticated.

\end{enumerate}

We will show that the following restricted asymptotic security condition holds. Under the above assumptions and conditioned on the complete publicly announced final detection pattern, the modified BB84 protocol is secure in the asymptotic limit against an arbitrary quantum attack by Eve on the surviving photonic system, whenever the observed key rate
\begin{equation}
\label{eq:modifiedAsymptoticKeyRate}
R(g) = \max\{0,p_s(g)p_p(1 - 2 H(\epsilon(g)))\},
\end{equation}
is strictly positive, where $\epsilon(g)$ is defined in \cref{eq:modifiedErrorRate} and $p_s(g)$ is defined in \cref{eq:modifiedSiftingProbability}. Here security means that, for any security parameters $s>0$ and $l>0$ chosen by Alice and Bob and for any eavesdropping strategy, the protocol either aborts or succeeds with probability at least $1-O(2^{-s})$. In the latter case, Eve's mutual information with the final uniformly random key is bounded above by $2^{-l}$.

We provide a sketch of the proof. We follow the entanglement-based reduction of BB84 in \cite{Shor00}. The entanglement-based version of the modified protocol is the same as that of standard BB84, except for an additional sifting step in which rounds corresponding to the erasure outcome $\ket{e}$ are discarded. The comparison procedure and the security parameters are otherwise the same as those of the modified prepare-and-measure protocol.

Conditioned on the complete final detection pattern, the surviving photonic systems belong to the logical qubit space $\mathcal{H}_L\oplus\text{span}\{\ket{e}\}$. Eve may perform an arbitrary quantum attack on these systems. The entanglement-based protocol can then be reduced to a CSS-code protocol in the same manner as in the standard BB84 security proof, with the additional sifting step used to exclude the erasure outcomes.

As detailed in the main text, the logical error rates entering the CSS-code analysis are therefore evaluated conditionally on the non-erasure events. The overall key rate must also include the additional survival probability associated with the erasure sifting. Combining the conditional CSS-code key fraction with this survival probability gives the asymptotic key rate in \cref{eq:modifiedAsymptoticKeyRate}. Hence, the protocol is asymptotically secure within the restricted setting whenever $R(g)>0$.

The security statement is restricted since the atom-field interaction is assumed to be trusted, fixed, memoryless, and inaccessible to Eve.


\bibliography{apssamp}

\end{document}


\title{Effect of atomic interactions on quantum communication protocols: supplemental document}


\author{Zixiang Hong}
\email{tom.hong@yale.edu}
\affiliation{Department of Applied and Computational Mathematics, Yale University, New Haven, CT, USA}
 
\author{John C. Schotland}
\email{john.schotland@yale.edu}
\affiliation{%
Department of Mathematics and Department of Physics, Yale University, New Haven, Connecticut 06511, USA
}%

\maketitle
\section{Equations of Motion}
In this section, we derive the equations of motion governing the dynamics of our model under real-space quantization. We consider a single photon interacting with an ensemble of atoms with constant dipole moments. The model is described by the Hamiltonian $H = H_F + H_A + H_I$, where the Hamiltonian of the field is given by \cref{eq:photonicHamiltonian}, the atomic Hamiltonian $H_A$ is given by \cref{eq:atomicHamiltonian}, and the interaction Hamiltonian is given by \cref{eq:interactionHamiltonian} in the paper. We restrict the dynamics to the single-excitation state
\begin{equation}
\ket{\Psi(t)}=\int d^3x\left[\psi(\bx,t)\cdot\boldsymbol{\hp}^\dagger(\bx)+n_0a(\bx,t)\hs^\dagger(\bx)\right]\ket{\emptyset},
\end{equation}
where $\psi(\bx,t)$ denotes the probability amplitude for finding the photon at position $\bx$ and time $t$, while $a(\bx,t)$ denotes the probability amplitude for exciting an atom at position $\bx$ and time $t$. The dynamics follow from the Schrödinger equation $i\frac{\partial}{\partial t} \ket{\Psi(t)} = H \ket{\Psi(t)}$. Projecting both sides onto the states $\boldsymbol{\hp}^\dagger(\bx) \ket{\emptyset}$ and $\sigma^\dagger(\bx) \ket{\emptyset}$ yields the equations of motion
\begin{equation}
\label{eq:EquationOfMotion1}
i\frac{\partial}{\partial t} \psi(\bx, t) = c(-\Delta)^{1/2} \psi(\bx, t)  + g^\ast P_{T}(n a\hat{\boldsymbol{d}})(\bx,t),
\end{equation}
\begin{equation}
\label{eq:EquationOfMotion2}
i\frac{\partial}{\partial t} a(\bx, t) = \Omega a(\bx, t) + g \psi(\bx, t) \cdot \hat{\boldsymbol{d}}.
\end{equation}

\section{Spontaneous Emission}
To validate our model, we derive the standard spontaneous emission rate. We consider a single atom located at the origin, corresponding to the atomic number density $n(\bx) = \delta(\bx)$. We further assume the initial conditions $\psi(\bx,0) = 0$ and $\abs{a(\0,0)}^2 = 1$. We adopt the following convention for the Fourier transform
\begin{equation}
\Tilde{f}(\bk) = \int d^3x \ e^{-i\bx\cdot \bk} f(\bx),
\end{equation}
and for the Laplace transform
\begin{equation}
f(z) = \int_0^\infty dt \ f(t) e^{-zt}.
\end{equation}
To solve the equations of motion, we Fourier transform \cref{eq:EquationOfMotion1} with respect to $\bx$ and Laplace transform it with respect to $t$. For \cref{eq:EquationOfMotion2}, we perform only the Laplace transform with respect to $t$ and evaluate the resulting equation at $\bx = \0$. This yields 
\begin{equation}
c\abs{\bk} \Tilde{\psi}(\bk, z) + g^\ast a(\0,z) (I- \hat \bk \otimes \hat \bk) \hat{\boldsymbol{d}} = i z \Tilde{\psi}(\bk, z),
\end{equation}
\begin{equation}
\Omega a(\0,z) + g \psi(\0,z) \cdot \hat{\boldsymbol{d}}= i[z a(\0,z) - a(\0,0)].
\end{equation}
Using the relation
\begin{equation}
\psi(\0,z) = \int \frac{d^3 k}{(2\pi)^3} \Tilde{\psi}(\bk,z),
\end{equation}
we eliminate $\Tilde{\psi}(\bk,z)$ from the above system and obtain
\begin{equation}
a(\0,z) = \frac{a(\0,0)}{iz - \Omega + \Sigma(z)},
\end{equation}
where 
\begin{equation}
\Sigma(z) = \abs{g}^2 \int \frac{d^3k}{(2\pi)^3} \frac{\hat{\boldsymbol{d}}\cdot (I - \hat{\boldsymbol{k}} \otimes \hat{\boldsymbol{k}} ) \hat{\boldsymbol{d}}}{c\abs{\bk} - i z - i \epsilon},
\end{equation}
is the self-energy. Within the pole approximation, we define the decay rate as
\begin{equation}
\Gamma = 2 \Im \Sigma(-i\Omega),
\end{equation}
and the Lamb shift as
\begin{equation}
\delta\omega = \Re \Sigma(-i\Omega),
\end{equation}
since the inverse Laplace transform gives
\begin{equation}
a(\0,t) = e^{-i(\Omega - \delta\omega) t} e^{-\frac{\Gamma}{2}t}.
\end{equation}
The probability of finding the atom in the excited state is therefore
\begin{equation}
\abs{a(\0,t)}^2 = e^{-\Gamma t}.
\end{equation}
Using the identity
\begin{equation}
\frac{1}{c\abs{\bk} - \Omega - i \epsilon} = p.v. \frac{1}{c\abs{\bk} - \Omega} + i \pi\delta(c\abs{\bk} - \Omega),
\end{equation}
where $p.v.$ denotes the principal value, we obtain
\begin{equation}
\Gamma = \frac{2\abs{g}^2 \Omega^2}{3\pi c^3}.
\end{equation}
Finally, using the exact expression for the coupling strength given in \cref{eq:couplingStrength}, we find
\begin{equation}
\Gamma = \frac{4\abs{\boldsymbol{d}}^2 \Omega^3}{3 c^3\hbar},
\end{equation}
which reproduces the standard result for the spontaneous emission rate. We note that this result with Markovian approximation agrees with the calculation retaining the full momentum dependence of $g_{\bk}$. Indeed, within the pole approximation, the emission rate receives contributions only from resonant modes satisfying $c\abs{\bk} = \Omega$, so replacing $g_{\bk}$ by its resonant value does not modify $\Gamma$. In contrast, this Markovian approximation modifies the Lamb shift because the corresponding principal-value integral also receives contributions from off-resonant modes.

\section{Constant Density Dynamics}
In this section, we consider a single photon interacting with a uniform-density ensemble of atoms with constant dipole moments, corresponding to $n(\bx) = n_0$. To solve the equations of motion in \cref{eq:EquationOfMotion1,eq:EquationOfMotion2}, we Fourier transform both equations with respect to $\bx$ and Laplace transform them with respect to $t$, yielding
\begin{equation}
\begin{split}
&i\left(z\tilde{\psi}(\bk,z)-\tilde{\psi}(\bk,t=0)\right)=c\abs{\bk}\tilde{\psi}(\bk,z) +n_0g^*\tilde{a}(\bk,z)
\left(I-\hat{\boldsymbol{k}}\otimes\hat{\boldsymbol{k}}\right)\hat{\boldsymbol{d}} ,\\
\end{split}
\end{equation}
\begin{equation}
\begin{split}
i\left(z\tilde{a}(\bk,z)-\tilde{a}(\bk,t=0)\right)&=\Omega\tilde{a}(\bk,z)+ g \tilde{\psi}(\bk,z)\cdot\hat{\boldsymbol{d}} ,
\end{split}
\end{equation}
where the initial conditions $\tilde{\psi}(\bk,t=0)$ and $\tilde{a}(\bk,t=0)$ are assumed to be known.

We now suppose that the excitation is initially in the field, so that $\tilde{a}(\bk,t=0) = 0$. Eliminating $\tilde{a}(\bk,z)$ from the coupled system gives an equation involving only the photon probability amplitude $\Tilde{\psi}$,
\begin{equation}
\begin{split}
\tilde{\psi}(\bk,z)
&=\frac{i}{iz-c\abs{\bk}}\tilde{\psi}(\bk,t=0)+\frac{n_0\abs{g}^2}{iz - c \abs{\bk}}\frac{\Tilde{\psi}(\bk, z) \cdot \hat{\boldsymbol{d}}}{iz-\Omega} (I-\hat{\bk}\otimes\hat{\bk})\hat{\boldsymbol{d}}.
\end{split}
\end{equation}
which can be rearranged as
\begin{equation}
\tilde{\psi}(\bk,z)
=\left(iz-c\abs{\bk}-\Sigma\right)^{-1} i\tilde{\psi}(\bk,t=0),
\end{equation}
where the operator $\Sigma$ is defined as
\begin{equation}
\begin{split}
\Sigma\tilde{\psi}(\bk,z)
&=\frac{n_0\abs{g}^2 }{iz-\Omega} \boldsymbol{d}_T(\bk)\cdot\Tilde{\psi}(\bk,z)\boldsymbol{d}_T(\bk),
\end{split}
\end{equation}
where $\boldsymbol{d}_T(\bk) = (I-\hat{\bk}\otimes\hat{\bk})\hat{\boldsymbol{d}}$.
Using the properties of the projector, we can evaluate the inverse of this operator explicitly, gives
\begin{equation}
\begin{split}
\tilde{\psi}(\bk,z)
&=\frac{i}{iz-c\abs{\bk}} (I-\hat{\boldsymbol{d}}_T(\bk)\otimes\hat{\boldsymbol{d}}_T(\bk) ) \tilde{\psi}(\bk,t=0) \\
&+\frac{-i(iz-\Omega)}{(iz-c\abs{\bk})(iz-\Omega)-n_0\abs{g}^2\abs{\boldsymbol{d}_T(\bk)}^2}\hat{\boldsymbol{d}}_T(\bk)\otimes\hat{\boldsymbol{d}}_T(\bk) \tilde{\psi}(\bk,t=0).
\end{split}
\end{equation}
Finally, after performing the inverse Laplace transform, we obtain the spectral wave function
\begin{equation}
\begin{split}
\tilde{\psi}(\bk,t)&=e^{-ic\abs{\bk}t}(I-\hat{\boldsymbol{d}}_T(\bk)\otimes\hat{\boldsymbol{d}}_T(\bk) ) \tilde{\psi}(\bk,t=0)\\
&+e^{-\frac{1}{2}it(C+c \abs{\bk} +\Omega)} \ \frac{ (C(1+e^{Ci t}) + (1 - e^{Ci t})(c\abs{\bk} - \Omega) )}{2 C}\hat{\boldsymbol{d}}_T(\bk)\otimes\hat{\boldsymbol{d}}_T(\bk) \tilde{\psi}(\bk,t=0),
\end{split}
\end{equation}
where $C = \sqrt{4n_0\abs{g}^2\abs{\boldsymbol{d}_T(\bk)}^2 +(\Omega-c\abs{\bk})^2} $.

